\documentclass[twoside]{article}
\usepackage{a4wide}
\usepackage{latexsym}
\usepackage{amsmath}
\usepackage{amsfonts}
\usepackage{amssymb}
\usepackage{times}
\usepackage{pstricks}
\usepackage{pst-node}
\psset{framearc=0.0,framesep=0.0truecm,nodesep=3pt,unit=0.75truecm,%
arrowsize=4pt 2,arrowinset=0.4}

\newcounter{mycnt}
\def\themycnt{\thesection.\arabic{mycnt}}
\def\mybenv#1{\refstepcounter{mycnt}%
       \vskip 3pt\noindent{\bf \themycnt~~#1}:~}
\def\myeenv{\hfill\rule{1ex}{1ex}\vskip 3pt}
\def\qed{\hfill$\Box$}
\def\ZP{{\mathbb{Z}_{+}}}
\def\ZT{{\mathbb{Z}_{2}}}
\def\DP{{\Delta^{\!+}}}
\def\etap{\eta^{+}}
\def\ep{{\epsilon^{+}}}
\def\up{{u^{+}}}
\def\sp{{\textsf{S}^{+}}}
\def\ot{{\,\otimes_{2}\,}}
\def\ds{{\,\oplus\,}}
\def\succ{\sigma}
\def\conv{\star}
\def\PS#1{[\kern-0.2ex[#1]\kern-0.2ex]}
\def\bfT{\textbf{T}}
\def\bfR{\textbf{R}}
\def\divp{\,\textrm{Div}}
\def\DM{{\Delta^{\!\cdot}}}
\def\DH{{\hat{\Delta}}}
\def\etam{\eta^{\cdot}}
\def\epm{{\epsilon^{\cdot}}}
\def\um{{u^{\cdot}}}
\def\sm{{\textsf{S}^{\cdot}}}
\def\DR{{\underline{\Delta}^{\!\cdot}}}
\def\DPR{{\underline{\Delta}^{\!+}}}
\def\sr{{\textsf{S}^\textsf{u}}}
\def\gcd{\,\textrm{gcd}}
\def\JJ#1{\mathop{\rule{1ex}{0.4pt}\rule{0.4pt}{1ex}_{#1}}} 
\def\eval{\,\textrm{eval}}
\def\Lambert{\,\textrm{Lambert}}
\def\Li{\,\textrm{Li}}
\def\odg{\,\textrm{odg}}
\def\add{\,\textrm{add}}
\def\pprime{{\prime\prime}}
\def\Id{\text{I\kern -0.27ex d}}
\def\dt{\frac{\text{d}}{\text{d\kern -0.1ex t}}}
\def\nn{\nonumber \\}
\def\pleth{\,\textrm{Pleth}}
\def\tens{\,\textrm{Tens}}
\def\sym{\,\textrm{Sym}}

\newcommand{\openk}{\Bbbk}
\newcommand{\openA}{\mathbb{A}}
\newcommand{\openB}{\mathbb{B}}

\newcommand{\openZ}{\mathbb{Z}}
\newcommand{\openR}{\mathbb{R}}
\newcommand{\openC}{\mathbb{C}}

\begin{document}

\title{The Dirichlet Hopf algebra of arithmetics} 
\author{%
Bertfried Fauser\thanks{%
Max Planck Institut for Mathematics in the Sciences, 
Inselstrasse 22-26, D-04103 Leipzig, Germany,
E-mail: fauser@mis.mpg.de}
\and
P.D. Jarvis\thanks{%
University of Tasmania,
School of Mathematics and Physics,
GPO Box 252-21, 7001 Hobart, TAS, Australia,
E-mail: Peter.Jarvis@utas.edu.au}
}
\date{November 28, 2005}

\maketitle
\begin{abstract}
Many constructs in mathematical physics entail notational complexities, deriving
from the manipulation of various types of index sets which often can be reduced
to labelling by various multisets of integers. In this work, we develop
systematically the ``Dirichlet Hopf algebra of arithmetics''  by dualizing the
addition and multiplication maps. Then we study the additive and multiplicative
antipodal convolutions which \textbf{fail} to give rise to Hopf algebra
structures, but form only a weaker Hopf gebra obeying a weakened homomorphism
axiom. A careful identification of the algebraic structures involved is done
featuring subtraction, division and derivations derived from coproducts and
chochains using branching operators. The consequences of the weakened structure
of a Hopf gebra on cohomology are explored, showing this has major impact on
number theory. This features multiplicativity versus complete multiplicativity of
number  theoretic arithmetic functions. The deficiency of not being a Hopf
algebra  is then cured by introducing an `unrenormalized' coproduct and an
`unrenormalized' pairing. It is then argued that exactly the failure of the
homomorphism  property (complete multiplicativity) for non-coprime integers is a
blueprint for the problems in quantum field theory (QFT) leading to the need
for renormalization. Renormalization turns out to be the morphism from the 
algebraically sound \textbf{Hopf algebra} to the physical and number 
theoretically meaningful \textbf{Hopf gebra} (literally: antipodal convolution). This
can be modelled alternatively by employing Rota-Baxter operators. We  stress the
need for a characteristic-free development where possible, to have a sound
starting point for generalizations of the algebraic structures. The last section
provides three key applications: symmetric function theory, quantum (matrix)
mechanics, and the combinatorics of renormalization in QFT which can be discerned
as functorially inherited from the development at the number-theoretic level as
outlined here. Hence the occurrence of number theoretic functions in QFT becomes
natural. 
\\
AMS Subject Classifications 2000: 
{\bf 
16W30, 
81T15, 
11N99, 
11M06  
} 
\end{abstract}
\eject

{\small\tableofcontents}

\section{Motivation}
\setcounter{equation}{0}\setcounter{mycnt}{0}

Many problems of mathematical physics are tackled by entities which are indexed
by (sub)sets of the nonnegative integers or sequences of nonnegative integers and
Cartesian sets formed out of them. Among them are found such  important cases as
the partitions of integers, appearing in group representations and symmetric
functions, or the occupation numbers of states of a quantum system, which also
involves representation theory. All these  plentiful examples come with an
additional structure when the objects are manipulated, for example using formal
power series. This can be translated in many cases into combinatorial properties
of the index sets, and sometimes into some arithmetic on them. It is then a
natural question to ask what kind of algebraic structure comes with the indices
of such objects as generating functions. 

It turns out that we are able to develop a coefficient-based approach directly
on the index sets. Hence we are dealing directly with the  arithmetic on the
nonnegative integers $\ZP$. It does not seem to be widely appreciated amongst
mathematical physicists or even some number theorists, that it is possible to
define additive and multiplicative comonoid structures on $\ZP$, which are dual
to addition and multiplication in $\ZP$. In some sense the question here is, in
which way a nonnegative integer can be decomposed additively or multiplicatively.
Such structures are of course not new -- for example the fact, that the
non-negative integers admit a partial ordering by divisibility is well-studied
via coalgebras on posets \cite{joni:rota:1979a}.

What us interests here is to exhibit clearly the fact that many standard
constructions in mathematics and mathematical physics are actually based on
additive and multiplicative comonoid constructions without these being made 
explicit. Our tool will be that of Hopf algebras, and we try to exploit this
tool as far as possible in a first exposition. While the additive structure
is not so obvious, the multiplicative one is very deeply involved in number
theory, combinatorics, representation theory, and last but not least in 
the renormalization theory of quantum fields.

We call both operations of addition and multiplication (plus, times)  $+$, 
$\cdot:\ZP\times\ZP\rightarrow \ZP$ `products', to make direct contact to  Hopf
algebraic notions. However, both pairs of `products'  and `coproducts', 
i.e.($+,\DP$) and ($\cdot,\DM$), \textbf{fail} to come up with a Hopf
\textbf{al}gebra structure, but form only a weaker  \textbf{Hopf gebra}, for
notation see below and in \cite{fauser:2002c}. That is, in both cases we find  an
antipodal convolution which allows to introduce \textit{some} crossing by
Oziewicz' theorem \cite{oziewicz:1997a} such that one comes up with a generalized
Hopf algebra fulfilling the homomorphism axiom for product and coproduct.
However, the natural crossing, the switch of two adjacent elements in the monoid
$\ZP^\ot$, does not allow the Hopf axioms to be fulfilled.

Our aim is to show, that the additive convolution and the multiplicative
convolution are intimately connected, as are addition and multiplication in a
ring. Furthermore, we want to show that the fact that the multiplicative
convolution, which we call \textbf{Dirichlet Hopf algebra} in a double abuse of
language, fails to be Hopf, having a deep number-theoretic counterpart.
Furthermore, we shall try to give arguments, that the  structure of
renormalization of quantum fields has exactly the same root. This might be one
step towards an explanation of why and how number theoretic functions appear
inevitably in renormalization, among them multiple zeta values.  

There are quite a few loose ends. The further iteration of multiplication to form
the noncommutative operation of exponentiation is not treated here. The present
structure is more compatible with a 2-category picture, but we refrained here
from exploring this in a first exposition. The cohomological considerations are
only taken up superficially, and a  much deeper study is needed to classify linear
forms etc. Topological issues of generating functions, i.e. convergence, have
been totally  neglected. This leaves us with formal results, which prevent for
the moment the use of our methods in analytic number theory where they are of 
interest. Further comments on these connections are given at the end of the
paper.

The paper is plainly structured so as to investigate first the additive
convolution, then the multiplicative convolution, the `Dirichlet Hopf algebra'.
The treatment of the Hopf convolutions is kept at a formal level, in order to
exhibit as much as possible the parallels between the additive and multiplicative
cases, and the introduction of associated constructions such as generating
functions and series is postponed until the Hopf algebraic details have been
discussed. Finally the interplay between the various structures is explored and
the unrenormalized coproduct is introduced. {\bf Renormalization} is the 
morphism which maps the Hopf algebra onto the antipodal convolution. The
first of these structures is necessary to be able to compute expansion formulae
so typical for perturbative QFT (pQFT). The last section provides three key
applications of the structure which may exhibit its ubiquitous appearance 
and importance:
\\
{\bf a:} {\bf Symmetric functions} provide the example where Rota and Stein
implicitly introduced much of the presently discussed material. There the
iterated structure of a plethystic Hopf algebra appears which is based on a
2-vectorspace $\tens \tens[ V]^+$.
\\
{\bf b:} It is demonstrated that the {\bf normal ordering} of quantum 
mechanical creation and annihilation operators, which produce the Stirling
numbers, can be modelled by the Dirichlet structure 
\textit{including renormalization}. This can be achieved in a less general 
setting by the usage of a Rota-Baxter operator. We demonstrate, that the
correct identification of algebraic structures is best done in a
characteristic free setting.
\\
{\bf c:} The example which coined the naming `renormalized' for some structures
is the application in {\bf renormalization theory of quantum  fields}. We show
how to make contact to the seminal work of Brouder and  Schmitt
\cite{brouder:schmitt:2002a} and thereby to Epstein-Glaser  renormalization
\cite{epstein:glaser:1973a}. We also discuss the connection of this combinatorial
approach to the Connes-Kreimer-BPHZ formalism of renormalization 
\cite{kreimer:1998a,connes:kreimer:2000a}.

\section{The Hopf convolution of addition}
\setcounter{equation}{0}\setcounter{mycnt}{0}

\subsection{Dualizing addition}

We start by considering the monoid of addition on nonnegative integers 
$\ZP$. The addition map is defined as the common addition of integers
\mybenv{Definition}
The commutative \textbf{addition} $+$ of nonnegative integers $\ZP$
is defined as usual
\begin{align}
+ &: \ZP \times \ZP \rightarrow \ZP \nn
n \times m &\mapsto +(n,m)=+(m,n)=n+m
\end{align}
\myeenv
Our aim is to dualize this addition, seen as a `product' on the 
number monoid, and to come up with a `coproduct' related to this 
addition. This, however, depends on the chosen duality. We will 
investigate two such choices in the sequel. We distinguish loosely
two types of tensor products: $\ot$ is linear over $\ZT = \ZP/2\ZP$,
and $\otimes$ is linear over $\openZ$. There arises some peculiarity, 
since we should carefully separate the scalars $\ZT$ or $\ZP$ from
the monoid elements $\ZP$, but that will cause no confusion here.
\mybenv{Definition}\label{KroneckerDual}
The \textbf{renormalized Kronecker duality} $K$ and the 
\textbf{renormalized Kronecker pairing}, also denoted $K$, are defined
as
\begin{align}
K : \ZP 
&\rightarrow \ZP^* \cong \hom(\ZP,\ZT) \nn
K(\ZP)(\ZP) 
&= \eval (\ZP^* \ot \ZP)  \nn
&= K(\ZP,\ZP) = \langle \ZP,\ZP\rangle \nn
n \times m &\mapsto \langle n\mid m \rangle = n^*(m)= K(n,m)=\delta_{n,m} 
\end{align}
where $\delta_{n,m}$ is the usual Kronecker delta.
\myeenv
This is our first choice, which might be the most natural to think of at 
a first glance if coming from numbers. Later it will become clear that the
second choice may be natural from a physical modelling point of view. The 
naming is chosen such that the adjectives `renormalized' and `unrenormalized'
fit the the usage in physics. We will have need to introduce an `unrenormalized'
pairing later, see theorem \ref{renKroneckerP}, which give a second duality.
\mybenv{Definition}\label{renKroneckerDual}
The \textbf{unrenormalized Kronecker duality} $R$ and the \textbf{unrenormalized
Kronecker pairing}, also denoted $R$, is defined as
\begin{align}
R : \ZP 
&\rightarrow \ZP^{\#} \cong \hom(\ZP,\ZP) \nn
R(\ZP)(\ZP) 
&= \eval (\ZP^{\#} \otimes_{\openZ} \ZP) \nn
&= R(\ZP,\ZP) = (\ZP,\ZP) \nn
n \times m &\mapsto (n\mid m) = n^{\#}(m)= R(n,m)
\end{align}
where we have introduced the dual $\ZP^{\#}$ with a different duality
that should not be confused with $\ZP^*$. 
\myeenv
Both \textit{Kronecker pairings} allow us to define duals of 
addition, the additive coproducts. We still use this terminology despite 
the fact that we deal with an additive structure. The `tensor monoid' is 
written using the more natural direct sum symbol $\ds$ in the additive
case. We will have need to carefully distinguish between \textit{units}, 
\textit{zeros} and \textit{ones}. 
\mybenv{Definition}
The \textbf{unit of addition} is the zero $0$. The injection of zero
into the algebra is $\etap : \ZT \rightarrow \ZP$ where $eval(\etap) = 0$. 
It will be sometimes crucial to distinguish the two notions $\etap\in\ZP$ and 
$0\in\ZT$ and we will use $\etap$ to denote $0$ injected into $\ZP$. 
\myeenv
This is evident from $0+n=n=n+0$. Using renormalized Kronecker duality, we get
\mybenv{Theorem}\label{DP}
The \textbf{renormalized Kronecker addition coproduct} $\DP$, with respect to 
$K$, is given as
\begin{align}
\DP(n) &= \sum_{r=0}^n r \ds (n-r) \nn
       &= n_{(1)} \ds n_{(2)} 
\end{align}
where the last line gives the notation using Sweedler indices
\cite{sweedler:1969a}. The addition coproduct $\DP$ is cocommutative 
and coassociative. 
\myeenv
\noindent
{\bf Proof:} We need to show that the coproduct $\DP$ is
Kronecker dual to addition. Therefore we consider
\begin{align}
\delta_{n,s+t} 
&= \langle n \mid s+t \rangle \nn
&= \langle \DP(n) \mid s\ds t\rangle \nn
&= \langle n_{(1)}\mid s\rangle \langle n_{(2)}\mid t\rangle \nn
&= \delta_{n_{(1)},s}\delta_{n_{(2)},t} 
\quad \forall s,t\text{~such that~} s+t=n
\end{align}
From which the definition follows. Cocommutativity is obvious, 
coassociatitvity follows either by duality or by a direct
short computation.\qed

The additive case is notationally somewhat peculiar. Strictly speaking 
we could deal with the monoid structure $(\ZP,+)$ of the Abelian 
semigroup of addition. By a slight abuse of notation we do not distinguish 
Hopf ring and Hopf algebra. Moreover, since we employ scalar valued 
linear forms\footnote{A further peculiarity of the additive case is, that the
only `scalars' are the $0$ and $1$ and all linear forms just map all elements 
to $\ZT$.} we need to define an action of scalars on the monoid and 
furthermore we want to be able to use the additive monoidal structure
$\ds$ to be able to consider pairs $(n,m)$ and even $r$-tuples 
$(n_1,\ldots,n_r)$ of nonnegative numbers. The action of the scalars $\ZT$ is
multiplicative $o\cdot n=0$ and $1\cdot n=n$, turning the monoid $\ZP$ into a
$\ZT$-module. Therefore we agree to use the following notation
\begin{align}
\etap \ds n &\cong n \cong n \ds \etap  \nn
\eval(\etap \ds n) & = \eval(n\ds \etap) = n \nn
\eval(0 \ds n) &= \eval(n\ds 0) = 0
\end{align}
This defines an isomorphism between tuples having units $\etap$ with such 
tuples where the units are omitted. A tuple having a zero $0$ will be mapped to
$0$ under the evaluation. This is the analogy of $ 1\otimes V \cong V 
\cong V\otimes 1$ in the multiplicatively written case. If the $\etap$ is 
evaluated in the underlying trivial ring, it will be zero, the additive unit. 
\mybenv{Definition}
The renormalized \textbf{Kronecker \textit{proper cut} addition coproduct} 
$\DP^\prime$ is given by the nonzero terms of the coproduct $\DP$ only
\begin{align}
\DP^\prime(n) 
&= \sum_{r=1}^{n-1} r\ds (n-r) \nn
\DP(n) 
&= \etap \ds n + n \ds \etap + \DP^\prime(n)
\end{align}
\myeenv
Sums over proper cuts, hence omitting the terms involving \textit{units},
here the zero, will be denoted $\sum^\prime$ for short.
\mybenv{Definition}
The \textbf{counit} $\ep$ of the additive coproduct $\DP$ is given as
\begin{align}
\ep &: \ZP \rightarrow \ZT \nn
n &\mapsto \ep(n)=\delta_{0,n}
\end{align}
\myeenv
{\bf Proof:} We need to check the defining relation of a counit.
This reads
\begin{align}
(\ep \ds \Id)\DP(n)
&= (\Id \ds \ep)\DP(n) \nn
&= \ep(n_{(1)}) \ds n_{(2)} \nn 
&= \delta_{0,n_{(1)}} n_{(2)}
 = \sum \delta_{0,r} (n-r) \nn
&= n 
\end{align}
showing $\ep$ to be a left and right counit.\qed

Note that the `scalars' $\ZT$ act multiplicatively on $r$-tuples 
$n_1\ds\ldots\ds n_r$. This is necessary to define the notion of a `linear map'
and to keep the analogy with the multiplicative case.

A further structural property of interest is that of primitiveness. An element $n$
of a comodule (comonoid) is called $(a,b)$-primitive, if it fulfils
\begin{align}
\DP(n) = n \ds a + b \ds n. 
\end{align}
Thus from the definition of $\DP$ we have the obvious

\mybenv{Theorem}
There is only one $(\etap,\etap)$-primitive element, called
primitive for short, the one $1$.
\myeenv

The deeper meaning of the theorem is that it allows the construction of 
the whole monoid $\ZP$ from this single primitive element $1$ and we could 
address $\ZP$ as the module so generated. This is well known from Peano 
axioms of natural numbers, where the successor map $\succ : \ZP \rightarrow \ZP$,
$\succ \mapsto n\rightarrow n+1$ allows the construction of all numbers out 
of zero by iteration: $1=\sigma(0)$, $2=\sigma(\sigma(0))$, and so on.

We note further that the counit is a homomorphism of addition as the
unit is an homomorphism of the coaddition. In terms of Hopf algebra 
theory this states that the monoid $\ZP$ and comonoid $\ZP$ are 
\textbf{connected}. In formulae this reads
\begin{align}
\ep(n+m) &= \ep(n)\ep(m) \in \ZT \nn
\DP(\etap) &= \etap \ds \etap
\hskip 1truecm (\DP(0)=0\ds 0)
\end{align}

\subsection{Antipodal convolution of addition}

From any pair of compatible, i.e. composable, product and coproduct
maps one is able to define a convolution for maps from the comonoid $B$ 
to the monoid $C$.
\mybenv{Definition}
A \textbf{convolution product} $\conv$ of maps $f,g : B \rightarrow C$
of a coproduct map $\Delta : A \rightarrow B\otimes B$ and a product
map $m : C\otimes C \rightarrow D$ is given as
\begin{align}
(f\conv g) : A &\rightarrow D \nn
(f\conv g)(a)  &= m (f \otimes g) \Delta(a) 
\end{align}
\myeenv
We will consider mainly endomaps $f,g : \ZP \rightarrow \ZP$ (or 
$\ZT \subset \ZP$) and complex valued maps $f,g : \ZP \rightarrow \openC$
where the injection $\iota : \ZP \rightarrow \openC$ is the canonical embedding
$\iota(n) = n + i 0 \in \openC$. The convolution is a map $* :
\hom(\ZP,\openC) \times \hom(\ZP,\openC) \rightarrow \hom(\ZP,\openC)$
turning this space into an algebra.

\mybenv{Theorem}
The (in general complex) convolution $(+,\DP)$ is 
unital with unit $\up = \etap\circ \ep$.
\myeenv
\noindent
{\bf Proof:} A trivial checking of the convolutional identity
\begin{align}
f \conv u = f = u \conv f
\end{align}
gives the result. The unit is unique due to biassociativity 
(associativity and coassociativity)\qed

\mybenv{Theorem}
The (in general complex) convolution $(+,\DP)$ is antipodal with 
antipode $\sp : \ZP \rightarrow \openC$, $\sp(n)=-n$.
\myeenv
\noindent
{\bf Proof:} The definition of an antipode is
\begin{align}
(\sp \ds \Id) = \up = (\Id \ds \sp)
\end{align}
Due to bicommutativity (cocommutativity and commutativity) we needed
to use only the first equation. Therefrom we get
\begin{align}
\sum \sp(r) \ds (n-r) &= \etap\circ\ep(n) = \delta_{0,n}\cdot\etap 
\end{align}
which can be solved recursively and yields $\sp(n)=-n$.\qed

The antipode does not exist if the codomain is $\ZP$, but needs
a codomain containing $\openZ$. If $\openZ$ is constructed from
pairs ($\ZP,\ZP$) modulo an equivalence relation, then the antipode
is realized as the switch of these pairs.

\mybenv{Theorem}\label{addOziewicz}
The convolution $(+,\DP)$ together with unit $\up$, counit $\ep$
and antipode $\sp$ does \textbf{not} form a Hopf algebra, but only
a Hopf gebra \cite{fauser:2002c}. 
\myeenv
{\bf Proof:} The compatibility axiom, using the switch 
$\textsf{sw} : \ZP\ot\ZP \rightarrow \ZP\ot\ZP$, turning a product-coproduct
map pair $(+,\DP)$ into a bialgebra maps fails to hold, as can be checked on
the element $2\ot 3$ providing a counterexample.

However, due to a theorem of Oziewicz 
\cite{oziewicz:1997a,oziewicz:2001b,fauser:oziewicz:2001a,fauser:2002c}, 
every antipodal convolution can be turned into a \textbf{Hopf gebra} 
with respect to the crossing
\begin{align}
(+ \ot +)(\Id \ds \DP \ds \Id)(\sp\ds \Id\ds \sp)(\Id\ds + \ds \Id)(\DP\ds \DP)
\end{align}
Note that the $+$ sign stands for the multilinear addition map and not 
literally for binary addition here. If the crossing is a braid the 
convolution will form a braided Hopf algebra and if the crossing 
is the (graded) switch one would be left with a (graded) Hopf algebra.
However, we know this fails to hold in our case, so that we remain in the
general Hopf gebra case.\qed

\mybenv{Theorem}
The antipode is an (anti)homomorphism of addition
\begin{align}
\sp(n+m) &= \sp(m)+ \sp(n) 
\end{align}
\myeenv
\noindent
{\bf Proof:} Trivial, due to the commutativity of addition the (anti)
does not make actual sense here. However, one should notice, that the
standard Hopf algebraic proof does not apply here, since we are not
dealing with a connected Hopf \textit{algebra} but only with a Hopf gebra, 
see also discussion in \cite{fauser:2002c,fauser:oziewicz:2001a}.\qed

\subsection{\label{sec:2-3}Branching operators and subtraction}

As studied in \cite{brouder:fauser:frabetti:oeckl:2002a} and
\cite{fauser:jarvis:2003a} we introduce \textbf{branching operators}.
Necessary ingredients for this are cochains and coboundary operators. 
An $n$-cochain is a $\ZT$-linear map from the $n$-tuple $\ZP^{\ds^n}$
into $\ZT$ (or $\openC$). A 1-cochain is hence a map $\ZP\rightarrow \ZT$. 
A particular 1-cochain is the counit $\ep$. 

The \textbf{convolutive inverse} $\phi^{-1}$ of a 1-chain $\phi$ is defined
such that $\phi^{-1}\conv\phi=\phi\conv\phi^{-1}=\ep$ that is
\begin{align}
+((\phi\ds\phi^{-1})\DP(n)) &= +((\phi^{-1}\ds\phi)\DP(n)) \,=\, \ep(n)\nn
\sum \phi^{-1}_r +\phi_{n-r} &= \delta_{n,0}\etap
\end{align}
This can be recursively solved. Consider an arbitrary series $\phi(i)=\phi_i$ 
forming a 1-cochain. The 
inverse reads 
\begin{align}
\phi^{-1}_{0} &= -\phi_{0},\quad
\phi^{-1}_{1} = -\phi_{1},\quad
\phi^{-1}_{2} = -\phi_{2}, \nn
\phi^{-1}_{3} &= -\phi_{3},\quad
\phi^{-1}_{4} = -\phi_{4},\quad
\phi^{-1}_{5} = -\phi_{5},\ldots
\end{align}
Hence one gets $\phi^{-1}_{i}=-\phi_i$. A normalized $1$-cochain is
defined to have $\phi_{0}=0$.

A coboundary operator may be defined along the lines given in the papers 
above using Sweedler cohomology \cite{sweedler:1968a}. We have
\begin{align}\label{coboundary}
\partial^i c_n(k_0,\ldots,k_n) &=\left\{
\begin{array}{cc}
\ep(k_0)c_n(k_1,\ldots,k_n) & i=0 \\
c_n(k_0,\ldots,k_{i-2},k_{i-1}\cdot k_i,k_{i+1},\ldots,k_n) & i\not=0,{n+1} \\
c_n(k_0,\ldots,k_{n-1})\ep(k_n) & i={n+1}
\end{array}\right. \nn
\partial_n c_n &= \partial_n^0 c_n \conv \partial_n^1 c_n^{-1} \conv
\ldots \conv\partial_n^n c_{n+1}^{\pm 1}=c_{n+1}
\end{align}
In the additive setting this spezializes as follows: a 1-cochain $\phi$ 
is a 1-cocycle if $(\partial_2\phi)(n,m)=0$, which is
equivalent to
\begin{align}
0&= (\partial_2\phi)(n,m)=
\sum_{r,s}(\ep(r)+\phi(n-r))-\phi(n+m)+(\phi(s)+\ep(m-s))
\end{align}
hence to an additive map
\begin{align}
\phi(n+m) &= \phi(n)+\phi(m)
\end{align}
A normalized 1-cocycle is then fully defined by the cocycle condition, 
and its value on $1$.
\begin{align}
\phi(0) &= 0, \quad \phi(1)=\phi_{1}\nn
\phi(n) &= n\cdot\phi_{1}\quad\text{since} \nn
\phi(n+m) &= \phi(n)+\phi(m) 
  = n\cdot\phi_{1}+m\cdot\phi_{1}
  = (n+m)\cdot\phi_{1}
\end{align}
An unnormalized 1-cochain cannot be a cocycle, since
$\phi(n)=\phi(n+0)=\phi(n)+\phi(0)\not=\phi(n)$ spoils the cocycle 
condition.

\mybenv{Definition}\label{boa}
A \textbf{branching operator} for the additive convolution is given by a 
1-cochain $\phi$ and the coproduct as
\begin{align}
/\Phi 
&= \eval((\phi \ds\Id)\DP) 
 = \eval((\Id \ds \phi)\DP) \nn
/\Phi(n) 
&=\phi(n_{(1)})\cdot n_{(2)} = \phi(n_{(2)})\cdot n_{(1)} 
\end{align} 
\myeenv

Note the asymmetry in this definition, using the multiplicative action of the
`scalars' $\ZT$ under evaluation. Special 1-cochains can be derived from the
renormalized Kronecker duality $\delta_{n,m}$. Let us introduce 
$\phi_b(n) = K(b)(n)=\delta_{b,n}$. In this way, via the evaluation map we can 
\textbf{curry} \cite{pierce:1991a} the $\ZT$-linear forms and parameterize them
by elements of $\ZP$. Hence we identified the dual $\ZP^*$ with $\ZP$ via the
Kronecker delta.

\mybenv{Theorem}
The branching operator $/\Phi_b$, $b\in \ZP$ fixed, with respect to the 
1-cochain $\phi_b$ acts as subtraction by $b$ if the argument is greater 
or equal to $b$, and as projection to $0$ otherwise.
\myeenv
\noindent
{\bf Proof:} We compute the branching operator as
\begin{align}
/\Phi_b(n) 
&= \eval((\phi_b \ot \Id)\DP(n)) \nn
&= \phi_b(n_{(1)})\cdot n_{(2)} = \sum_{r=0}^n \phi_b(r)\cdot (n-r) \nn
&= \sum \delta_{b,r}\cdot (n-r) = \left\{
\begin{array}{cl}
n-b & \text{if $n\ge b$} \\
0   & \text{otherwise}
\end{array}\right.
\end{align}
showing the desired feature.\qed

Note, that while the antipode $\sp$ was a map which necessarily enlarged
the codomain, the subtraction established by employing branching
operators here can still be established as an endomorphism of $\ZP$.

\subsection{\label{sec:2-4}Contractions and a derivation}

Branchings are related to contractions via an identification of $V$ with
$V^*$ using $K$ or $V^{\#}$ using $R$. While a linear form $\phi_b$ acts 
in a branching the `name' $b$ can be dualized and acts via a contraction 
$b\JJ{K} n= \eval(K(b)\ds n)$. In differential geometry one would write 
$\phi_b(x) \cong i_b(x)$ since it forms an inner derivation if $b$ is 
primitive.

Since we have only one primitive element $1$, the $(\etap,\etap)$-primitive
element to which we could assign the grade 1, in this convolution the 
$1$, we can expect only one \textbf{derivation} acting on the additive
convolution. Let us define $1^*=K(1)$ as the Kronecker dual linear form
$1^*(n)=\delta_{1,n}=K(1,n)$ and compute
\begin{align}
1^*(n) 
&=(\eval\ds \Id)(K\ds \Id)(\Id\ds \DP)(1 \ds n) \nn
&= \sum_r 1^*(r)\cdot (n-r) = \sum_r \delta_{1,r}\cdot (n-r) = n-1
\end{align}
Besides the fact that this is not literally a derivation, it cannot fulfil
automatically the Leibniz rule, since the homomorphism axiom does not hold
in this convolution. Actually a `derivation' of $n$ by $1$ would need to
produce as $n(n-1)$ due to the fact that there are $n$ possible ways to
extract a $1$ out of $n=1+\ldots+1$. Since we do not need it here, we
postpone the solution to this problem until we have studied the 
multiplicative case. However, if one thinks in terms of successor maps
a derivation would be a much more natural object to think of. It would
be the derivative w.r.t. the successor map.

\subsection{\label{sec:2-5}Ordinary polynomial series generating functions}

With the abstract notion developed so far, we want to give a first
application, which also sheds light on the potential field of usage of the
present ideas.

Let us consider an infinite sequence of nonnegative integers (integers 
in general) $a_1,a_2,\ldots,a_n,\ldots$. To cope with such data, one 
usually resorts to generating functions. We will use ordinary polynomial
series generating functions (opsgf, see \cite{wilf:1990a}). Hence 
we introduce a formal indeterminate, say $t$. We do \textit{not}
specify the domain of this variable (yet), but we demand that the
$a_n$ and $t$ commute, and that $t$ is at least power associative
and not nilpotent, hence that we can introduce arbitrary powers of 
$t$ recursively.

\mybenv{Definition}
An \textbf{ordinary polynomial series generating function} (opsgf)
is an element of the ring of formal power series $\ZP\PS{t}$. Addition
in this ring is component wise addition of power series, multiplication
is given by pointwise multiplication of power series.
\begin{align}\label{opsgf}
(f+g)(t) 
&=\sum f_nt^n + g_nt^n = \sum (f_n+g_n)t^n \nn
(f\cdot g)(t)
&=\sum_{n\ge 0} \sum_{m\ge 0} f_nt^n g_m t^m \nn 
&=\sum_{n\ge 0} \sum_{r= 0}^{n} (f_r\cdot g_{n-r}) t^n 
 =\sum h_nt^n 
\quad\text{Cauchy product formula}\nn
\text{hence~}\DP(n)
&= \sum r\ds (n-r)\quad \text{on the indices.}
\end{align}
\myeenv

A more formal way to look at this is to say ($\ds$ is our monoid product)
\begin{align}
(f+g)(t)      &= \sum_{n\ge0} +(f\ds g)\DP(n) t^n \nn
(f\cdot g)(t) &= \sum_{n\ge0} \cdot(f\ds g)\DP(n) t^n
\end{align}
This unveils the usage of the coproduct of addition in the Cauchy formula.
Note, that the process of forming a formal power series employs a 
duality. Consider the series $a_0,a_1,a_2,\ldots,a_n,\ldots$ i.e. an 
element of $\openZ_{+}^\infty$ and the series $1=t^0, t=t^1, t^2, \ldots,
t^n,\ldots$ which pair element wise to form an element of $\ZP\PS{t}$.
Note further, that multiplying powers of $t$, due to power associativity,
amounts to \textit{adding} the exponents, a basic fact used in 
combinatorics. Hence one finds here the addition of nonnegative integers.
The coefficients have, due to the Cauchy product formula, a quite
different law of composition, which we recognize immediately to be the
Kronecker coproduct of addition. In a certain sense these two paired series
are dual if evaluation is considered.

We may make this more explicit using `letter-place' techniques as often
employed by Rota. Let $\openA$ be an alphabet, i.e. the formal sum of
the $a_n$, which we take as letter. Further, let $\bfT$ be a formal
alphabet of the $t^n$, and use it as a `place'. The formal power series
emerges as a pairing between letters and places, 
$(\openA\mid \bfT)\in \ZP\PS{t}$. We may now restate the laws
governing the structure of the power series rings more formally as
\begin{align}
(\openA\mid\bfT) + (\openB\mid \bfT) 
&= (\openA+\openB\mid \bfT) 
 = (\openA\mid \bfT_{\{1\}})(\openB\mid \bfT_{\{2\}}) \nn
(\openA\mid\bfT)\cdot(\openB\mid\bfT)
&= (\openA\cdot\openB\mid\bfT) = 
   (\openA\mid \bfT_{(1)})(\openB\mid\bfT_{(2)}) \nn
\Delta_{\emptyset}(\bfT) = \bfT_{\{1\}}\otimes \bfT_{\{2\}}
&\Leftrightarrow   
\Delta_{\emptyset}(t^n) = t^n\otimes 1 + 1\otimes t^n \nn
&{~}~~~~~~ (\openA\mid 1)=1=(\openB\mid 1) \nn   
\Delta(\bfT)=\bfT_{(1)}\otimes \bfT_{(2)} 
&\Leftrightarrow 
\Delta(t^n) = \sum_{r\ge 0}^n t^r \otimes t^{n-r}
\end{align}
Hence we see that the antipodal Kronecker Hopf convolution of 
addition governs the ring structure of the formal power series
i.e. that of ordinary polynomial series generating functions.
This fact is not a coincidence when the categorial background is
taken into account; however a detailed development is beyond the
scope of this work.

Noting that the variable $t$ was not fixed for a special domain, we
could take $t$ as a placeholder for an irreducible representation
space of a group $t\equiv V^\lambda\cong \openC^{\vert\lambda\vert}
=\langle z_1,z_2,\ldots \rangle$, and $\lambda$ an integer partition. 
In this case, the powers of $t$ would have to be considered as symmetrized
powers of such a vector space $V^\lambda$. This leads to the well known 
fact that the algebra of formal power series is equivalent to the symmetric
algebra over a vector space over a ring {\bf R}
\begin{align}
\ZP\PS{t} 
&\cong \sym(V^\lambda) = {\bf R} \oplus V^\lambda \oplus 
\otimes^2 V^\lambda \oplus \ldots 
\end{align}
we will employ this in section (\ref{sec:SymFct}).

Finally we take up the opportunity to identify the dual notion
of this construction. Introducing $\partial_t$, the formal derivative
with respect to $t$ as a dual of $t$ we find the action
\begin{align}
\partial_t (t) &= 1 \nn
\frac{\partial^n_t}{n!} t^m \vert_{t=0} &=\delta_{n,m} 
\end{align}
Sometimes the procedure to extract the $n$-th coefficient of a power
series $f(t)$ is denoted by $[t^n]f(t)= f_n$, see \cite{wilf:1990a}
for notation. Hence one has $[t^n]= 1/n!\partial^n_t\vert_{t=0}$. 
However, this explicit construction is possible in characteristic 0
only. If one wants to avoid this complication, one sticks with the 
abstract notion of a new type of variable. These formal dual elements 
$[t^n]$ fulfil now a slightly different algebra
\begin{align}
[t^n]\cdot [t^m] &= {n+m\choose n} [t^{n+m}]
\end{align}
This is a \textbf{divided powers algebra} $\divp(\bfT)$  and the elements 
are usually written as $t^{(n)}$ with parentheses around the exponents.
We will use in the following $t^{(n)}$ and $t^{\# n}$ synonymously for 
$[t^n]$. One should note, that
a power series algebra based on divided powers has the same properties
as an exponential generating function. Indeed in characteristic zero
these notions are interchangeable. Once more it is now easy to see the
well known fact that the dual of a symmetric algebra in any characteristic
is given by a divided powers algebra (and not in general by another 
symmetric algebra, which is possible in characteristic zero only)
\begin{align}\label{SymDivp}
\sym(V)^{\#} &\cong \divp(V^{\#})
\end{align}
This is in full accord with experience, that derivations, i.e. duals of
variables, are exponentiated in group theory or Fourier analysis,
which is modelled here employing divided powers. Letter-place techniques
and divided powers were exactly employed in these areas for that 
particular reason, see e.g. 
\cite{akin:buchsbaum:weyman:1982a,akin:buchsbaum:1985a,akin:buchsbaum:1988a} 
and references therein.

We close this discussion by providing the divided powers algebra coproduct
of addition, which produces the same terms as (\ref{DP}) but with additional
binomial weightings. Let $f,g$ be formal power series in the dual variables
$t^{\#}$ which are divided powers, we find
\begin{align}\label{dpgf}
f(t^{\#}) \cdot g(t^{\#}) 
&= \sum_{n,m\ge 0} f_n g_m t^{\#n}t^{\#m} \nn
&= \sum_{n\ge 0} \sum_{r 0}^{n} f_r g_{n-r} {n \choose r} t^{\#n} \nn
&= \sum_{n\ge0}\cdot(f\ds g)\DPR(n)t^n
\end{align}
This gives the `unrenormalized coproduct of addition' as we will see
below in theorem (\ref{DR})
\begin{align}
\DPR(n) &= \sum_{r= 0}^{n} {n \choose r} r \ds (n-r) 
\end{align}
From a combinatorial point of view it is of great importance to know
where these coefficients come from to have a proper interpretation of 
the combinatorial meaning of series in such dual variables. An ingenious 
usage of these ideas may be found in \cite{rota:stein:1994a,rota:stein:1994b}.
For a discussion of polynomials associated with binomial coalgebras 
see \cite{rota:etal:1975a} and references therein.

\section{The Dirichlet Hopf algebra} 
\setcounter{equation}{0}\setcounter{mycnt}{0}

\subsection{Dualizing multiplication}

Since arithmetic comes with addition and multiplication, we proceed
to dualize multiplication along the same lines as we did with addition.
For the first step we once more use Kronecker duality, definition
\ref{KroneckerDual}, to achieve this.
\mybenv{Definition}
The commutative \textbf{multiplication} $\cdot$ of nonnegative
integers $\ZP$ is defined as
\begin{align}
\cdot &: \ZP\times \ZP \rightarrow \ZP \nn
n\times m &\mapsto \cdot(n,m)=\cdot(m,n)=n\cdot m
\end{align}
The \textbf{unit of multiplication} is the one $1$. The injection of 
the unit into the monoid $\ZP$ is denoted $\etam$
\myeenv

This is evident from $1\cdot n=n=n\cdot 1$. The product is commutative
and associative.

\mybenv{Theorem}\label{DM}
The \textbf{renormalized Kronecker multiplication coproduct} $\DM$ is 
given as
\begin{align}
\DM(n) &= \sum_{d\mid n} d \ot \frac{n}{d} \nn
       &= n_{[1]}\ot n_{[2]}
\end{align}
where $d\mid n$ denotes `$d$ divides $n$', the sum is over all divisors 
$d$, and we used the Brouder-Schmitt convention \cite{brouder:schmitt:2002a}
to indicate different coproducts by differently shaped Sweedler brackets.
\myeenv
\noindent
{\bf Proof:} We need to show that the coproduct $\DM$ is dual to
multiplication. Therefore we compute
\begin{align}
\delta_{n,s\cdot t} 
&= \langle n \mid s\cdot t\rangle \nn
&= \langle \DM(n) \mid s \ot t\rangle \nn
&= \langle n_{[1]} \mid s\rangle\langle n_{[2]} \mid t\rangle \nn
&= \delta_{n_{[1]},s}\delta_{n_{[2]},t}
\quad\forall s,t\text{~such that~} s\cdot t = n
\end{align}
From which the definition follows. Cocommutativity is obvious from the
construction, coassociativity follows by duality or from a short 
computation.\qed 

Computations involving this coproduct are costly, since the explicit
knowledge of all divisors of an integer involves its prime number
factorization. However, it will serve us as a formal device very well.

\mybenv{Definition}
The renormalized \textbf{Kronecker \textit{proper cut} multiplication 
coproduct} $\DM^\prime$ is given by the nontrivial terms of the coproduct 
only
\begin{align}
\DM^\prime(n) &= \sum_{d\mid n \atop d,n\not=\{1,n\}}^{\prime} 
d \ot \frac{n}{d}\nn
\DM(n)&= \etam\ot n + n \ot\etam + \DM^\prime(n)
\end{align}
We denote sums of proper cuts once more as $\sum^\prime_{d\mid n}$.
\myeenv

It is readily checked that $\frac{n}{d}$ is in $\ZP$ since $d$ is a divisor of
$n$ and the definition is meaningful. 

\mybenv{Definition}
The \textbf{counit} $\epm$ of the multiplication coproduct $\DM$ is given as 
\begin{align}
\epm &: \ZP \rightarrow \ZT \nn
n &\mapsto \epm(n)=\delta_{1,n}
\end{align}
\myeenv
\noindent
{\bf Proof:} For notational convenience, we drop the explicit display of
multiplication from now on. We need to check the defining relation of a 
counit. This reads
\begin{align}
(\epm \ot \Id)\DM(n)
&=(\Id\ot \epm)\DM(n) \nn
&=\epm(n_{[1]})\ot n_{[2]} \nn
&= \delta_{1,n_{[1]}} n_{[2]} = \sum_{d\mid n} \delta_{1,d} \frac{n}{d} \nn
&= n
\end{align}
showing $\epm$ to be a left and right counit. This counit is unique due to
biassociativity.\qed

\mybenv{Theorem}
The $(\etam,\etam)$-primitive elements are exactly the prime integers $p_i$
\myeenv
\noindent
{\bf Proof:} The unique prime number decomposition of an integer $n$ is given
as $n=p_1^{r_1}p_2^{r_2}\ldots p_k^{r_k}$. The divisors of $n$ can be formed
from similar expressions where the exponents $\{s_1,\ldots,s_k\}$ form a
sub multiset of the multiset $\{1^{r_1},\ldots,k^{r_k}\}$. The unit is 
written as $\etam=p_i^0$ for some $i$ and its coproduct has the form 
$\DM(\etam)=\etam\ot\etam$. Hence the unit is not a primitive element 
(and not a prime). Due to the fact that 
$\DM(\prod_i p_i^{r_i})= \prod_i \DM(p_i^{r_i})$, only such
sets $\{r_1,\ldots,r_k\}$ which have \textit{exactly} one $1$ and other
elements zero give rise to exactly two divisors, $n$ itself and the unit.
Hence we find
\begin{align}
\DM(p_i) &= p_i\ot \etam + \etam \ot p_i \qquad \forall p_i \in \text{~primes}
\end{align} 
as the only primitive elements.\qed

\mybenv{Theorem}
The monoid $\ZP^\ot$ is connected w.r.t. multiplication and comultiplication.
\myeenv
\noindent
{\bf Proof:} We need to check
\begin{align}
\epm(n\cdot m) &= \epm(n)\epm(m) \nn
\DM(\etam) &= \etam \ot \etam
\end{align}
the first is obvious, the second was discussed in the preceding proof.\qed

\subsection{The Dirichlet convolution of arithmetic functions}

We will proceed to establish an antipodal convolution as in the case of
addition. As in subsection \ref{sec:2-5} we will identify it with the
well studied convolution ring of (multiplicative) arithmetic functions. The
identification will become clear \textit{a posteriori}. We postpone
furthermore the precise definition of the complex functions 
$f(s)=\sum_{n\ge1}f_n n^{-s}$ attached to the series of complex numbers
(integers) $f_1,f_2,\ldots ,f_n,\ldots$ until we discuss these generating
functions in section \ref{sec:DGF}. However we use the term `function'
for such series from now on by abuse of language. Note that $f(n)=f_n$ is
a series element but $f(s)$, $s\in \openC$ will be a complex valued 
function.

We consider series of complex numbers (integers) $f_1,f_2,\ldots,f_n,\ldots$
and therewith related functions $f,g,\ldots : \ZP \rightarrow \openC$. These
functions are called arithmetic functions in number theory, see e.g.
\cite{bruedern:1995a}. The addition of these functions is defined term wise
\begin{align}
(f+g)(n) &= f(n)+g(n) = f_n+g_n
\end{align}
where the multiplication is given by Dirichlet convolution, which we connect
with multiplication and the Kronecker multiplication coproduct.

\mybenv{Definition}
The \textbf{Hadamard product} $. : f\times g \mapsto f.g$ of arithmetic 
functions is given as the \textit{term wise} multiplication
\begin{align}
(f.g)(n) &= f(n)g(n) = f_n g_n
\end{align}
\myeenv

Actually this gives the series a product ring structure, which we will need
below. We can define a group like coproduct $\delta : \ZP \rightarrow \ZP
\ot \ZP$ in such a way that all elements just double $\delta(n)=n\ot
n$. The Hadamard product becomes then the convolution product of the pair
($\cdot,\delta$). These transformations play an important role in the the
theory of Schur-Weyl duality in the symmetric group \cite{scharf:thibon:1994a}.
One could make use of them in the present setting too, which we will
demonstrate elsewhere.

\mybenv{Definition}\label{DirichletConv}
The \textbf{Dirichlet convolution} $(\cdot,\DM)$ of maps $f,g : \ZP
\rightarrow \openC$ or $\ZP$ is defined as
\begin{align}
(f\conv g) &: \ZP \rightarrow \openC \nn
(f\conv g)(\ZP) &\mapsto 
(f\conv g)(n) = \cdot (f\otimes g)\DM(n)
              = \sum_{d\mid n} f(d)g(\frac{n}{d})
\end{align}
\myeenv
We will see in section \ref{sec:DGF} that this product is intimately related
with the the point wise product of Dirichlet generating functions.
\mybenv{Definition}
The \textbf{Dirichlet convolution ring} of arithmetic functions is defined
via component wise addition of the series and the Dirichlet convolution
as product.
\myeenv

\mybenv{Theorem}
The Dirichlet convolution is unital with unit $\um=\etam\circ \epm$ 
\myeenv
\noindent
{\bf Proof:} Since we know the product to be unital and the coproduct to be
counital, it is a matter of checking 
\begin{align}
f\conv u &= f = u\conv f
\end{align}
to see that the unit  $\um=\etam\circ\epm$ is a unit for the convolution.
This unit is unique due to biassociativity.\qed

\mybenv{Definition}
The M\"obius function $\mu(s)$ is given as Dirichlet series w.r.t. the formal
parameter $s$ (usually taken complex valued, definition below Sec. 
\ref{sec:DGF})
\begin{align}
\mu(s) &= \sum_{n>0} \frac{\mu_n}{n^{s}}
\end{align}
defining the series $\{\mu_n\} = \{1,-1,-1,0,-1,1,-1,0,0,1, \ldots \}$.
The coefficients are given by
\begin{align}
\mu_n &=\left\{
\begin{array}{cl}
1 & n=1 \\
(-1)^{k} & n=p_1^{r_1} p_2^{r_1} \ldots p_k^{r_k}\quad\text{where~}
  r_i\in\{0,1\}~\forall i \\
0 & \text{otherwise} 
\end{array}\right.
\end{align}
\myeenv

The M\"obius function is hence a (signed) projection onto square-free
integers where the sign is negative for an odd number of mutually distinct 
primes and positive for an even number of mutually distinct primes. By 
definition one has $\mu_1=1$.

\mybenv{Theorem}\label{antipm}
The Dirichlet convolution is antipodal with antipode $\sm= n\cdot\mu(n)$ 
where $\mu(n)$ is the (series of coefficients of the) M\"obius function.
Hence the antipode $\sm$ is given as the Hadamard product of the identity
times the M\"obius function $(\Id.\mu)(s)$, where 
$\Id(s)= \sum_{n\ge1}\Id_n/n^{s}$ and $\Id_n=n$, hence 
$\Id(s) = \sum_{n\ge1}1/n^{s-1}$, see identification as series
in eqn. (\ref{dirichletantip}).
\myeenv
\noindent
{\bf Proof:} We use the definition of the antipode to compute
\begin{align}
(\sm\conv \Id)(n) 
&= \um(n) = (\Id\conv \sm)(n) \nn
&= \sum_{[n]} \sm(n_{[1]}) n_{[2]}\nn
&= \sum_{d\mid n} \sm(d) \frac{n}{d}\nn
&= \etam\circ\epm(n) = \delta_{1,n}\etam   
\end{align}
This equation can be recursively solved yielding the result.\qed

The present result is well known in number theory as M\"obius 
inversion, see e.g. \cite{apostol:1979a}.
\mybenv{Examples}
\begin{align}
\sm(1)&=1,\quad \sm(2)=-2,\quad \sm(3)=-3, \nn
\sm(4)&= -\sm(2)2-\sm(1)4+0 = 0, \quad \sm(5)=-5,\nn
\sm(6)&= -\sm(3)2-\sm(2)3-\sm(1)6+0 = 6, \ldots 
\end{align}
\myeenv

Two numbers $n,m$ are called relatively prime, if their greatest common divisor
(gcd) is 1, denoted as $\gcd(n,m)=1$ or $(n\mid m)=1$ for short. 

\mybenv{Definition}
An arithmetic function $f : \ZP \rightarrow \openC$ is called
\textbf{multiplicative}, if for any two relative prime numbers $n,m$ 
\begin{align}
f(n\cdot m) &= f(n)f(m)
\end{align}
it is called \textbf{complete multiplicative} if this holds true for all 
pairs of numbers $n,m$.
\myeenv
Another way to say that a function is complete multiplicative is to state 
that it is a homomorphism of the multiplicative structure. Hence number 
theoretic functions \textit{fail} in general to be homomorphisms on products
of non relative prime integers $n,m$. Many important number theoretic
functions are multiplicative but \textit{not} totally multiplicative.
Examples of multiplicative functions are the Euler phi-function, the
Legendre symbol, the number of divisors of $n$ function, 

\mybenv{Theorem}
The Dirichlet convolution does \textbf{not} form a Hopf algebra but only a 
Hopf gebra. However, on relative prime numbers $n,m$ the compatibility axiom 
holds true. 
\myeenv

This means, that the product is a multiplicative map on relative
prime elements of the comonoid $\ZP$ and the coproduct $\DM$ is a
multiplicative map on relative prime elements of the monoid $\ZP$. 

\noindent
{\bf Proof:} We take $2,2$ with $\gcd(2,2)=2$ and compute (using percent 
signs to abbreviate parts of the formula)
\begin{align}
\DM(2\cdot 2) &= \sum_{d\mid 4} d \ot \frac{4}{d} 
= 1\ot 4 + 2\ot 2 + 4\ot 1 \nn
&~\hskip -2truecm
(\cdot\ot \cdot)(\Id\ot \textsf{sw}\ot \Id)(\DM\ot \DM)(2\ot 2) \nn
&= (\%)(\%)\left( (2,1,2,1)+(1,2,2,1)+(2,1,1,2)+(1,2,1,2)\right) \nn
&= (\%) \left( (2,2,1,1)+(1,2,2,1)+(2,1,1,2)+(1,1,2,2) \right) \nn
&= 4\ot 1 + 2\ot 2 + 2 \ot 2 + 1\ot 4  
\end{align} 
showing that there appears a multiplicity and this serves as a counter
example for total multiplicativity. However, a similar consideration for 
general coprime $n,m$ shows the identity to hold.\qed

We come now to a fact which has major implications on the applicability of
the cohomological considerations. This will have also great impact on the
potential usage and interpretation in physics.
\mybenv{Theorem}
A $1$-cochain $\phi$ is a $1$-cocycle in the Dirichlet convolution ring, 
if it is a complete multiplicative map, that is
\begin{align}
\phi(n\cdot m) &= \phi(n)\phi(m)
\end{align}
for all $n,m$ in $\ZP$.
\myeenv
\noindent
{\bf Proof:} The cocycle identity (\ref{coboundary}) reads 
\begin{align}
(\partial_2\phi)(n,m) &= \epm(n)\epm(m)
\end{align}
From this identity we derive using the convolution product
\begin{align}
\sum_{{d\vert n}\atop{s\vert m}}
  \phi(\frac{n}{d})\phi(\frac{m}{s})\phi^{-1}(d\cdot s) 
&= \epm(n)\epm(m) = \epm(n\cdot m) \nn
\sum_{{{d\vert n}\atop{s\vert m}}\atop{r\vert d\cdot s}}
\phi(\frac{n}{d})\phi(\frac{m}{s})\phi^{-1}(\frac{d\cdot s}{r})\phi(r)
&= 
\sum_{{r\vert d\cdot s}}
\epm(\frac{n\cdot m}{r})\phi(r) \nn
\sum_{{d\vert n}\atop{s\vert m}}
\phi(\frac{n}{d})\phi(\frac{m}{s})\epm(d\cdot s)
&= \phi(n\cdot m) \nn
\phi(n)\phi(m)&=\phi(n\cdot m) 
\end{align}
where we used $\epm(n\cdot m)=\delta_{n\cdot m,1} \Leftrightarrow n=1$, $m=1$.
This computation is valid for all $n,m$ since we did not make use of the 
homomorphism axiom, but the converse is not true for non-complete multiplicative 
$1$-cochains, see example below.\qed

\mybenv{Example}
To exhibit the peculiarities appearing for non-complete multiplicativity,
we give an example of multiplicative functions and their derived 2-cocycles
in the case of the M\"obius series and the zeta series. We write $\mu(n)$ 
for the $n$th element of the series related to the M\"obius function
$\mu(s)$. Note that the M\"obius series is not a homomorphism since 
$\mu(2)\mu(2)=-1\cdot -1=1$ but $\mu(2\cdot 2)=\mu(4)=0$. Let $\partial_2$ 
be the coboundary operator as introduced in (\ref{coboundary}), see 
\cite{sweedler:1968a}, and employed in 
\cite{brouder:fauser:frabetti:oeckl:2002a,fauser:jarvis:2003a}.
It maps 1-cochains into 2-coboundaries. The related M\"obius 2-cocycle is 
hence given, using $\mu\conv\zeta=\etam$, that is $\mu^{-1}=\zeta$, as
\begin{align}
(\partial\mu)(n,m) 
&= \sum_{{{d\mid n}\atop{l\mid m}}}
   \mu(d)\mu(l)\zeta(\frac{n}{d}\cdot\frac{m}{l}) \nn
&= \sum_{{{d\mid n}\atop{l\mid m}}}\mu(d)\mu(l) \nn
&=\sum_{d\mid n}\mu(d) \sum_{l\mid m}\mu(l) \nn
&=\epm(n)\epm(m)
\end{align}
Since $\sum_{d\mid n}\mu[d]=\delta_{n,1}=\epm(n)$. Thus the M\"obius 
$1$-cochain $\mu$, besides not being a homomorphism, nevertheless gives
rise to a trivial 2-cocycle, and turns out to be a 1-cocycle! A question
which has to be settled is thus if all multiplicative but not complete 
multiplicative functions give rise to trivial $2$-cocycles. However, the 
calculation for the convolutive inverse $1$-cochain $\zeta$ amazingly 
gives
\begin{align}
(\partial_2\zeta)(n,m) 
&= \sum_{{{d\mid n}\atop{l\mid m}}}
   \zeta(d)\zeta(l)\mu(\frac{n}{d}\cdot\frac{m}{l}) \nn
&= \sum_{{{d\mid n}\atop{l\mid m}}}
   \mu(\frac{n}{d}\cdot\frac{m}{l})
\end{align}
which is by no means in an obvious way trivial. If one computes the
characteristic polynomials of $n$ by $n$ upper left submatrices, one gets
quite nontrivial polynomials. The zeros of these polynomials are in $\openZ$
with certain multiplicities. It is hence false that all multiplicative 
functions form \textit{trivial} 2-cocycles. This sheds some light on how 
cohomology may help to deal with number theoretic questions. The seeming 
discrepancy between this example and the theorem is explained by the 
observation that for \textit{complete multiplicative} $\phi$ one finds
\begin{align}
(\partial_2\phi)^{-1} &= (\partial_2\phi^{-1})
\end{align}
The M\"obius function hence does not allow to the analogous computation, 
explaining the astonishing fact that
\begin{align}
(\partial_2\mu)^{-1} &\not= (\partial_2\mu^{-1}) = (\partial_2\zeta)
\end{align}
despite the fact that $\mu\conv\zeta=\epm$. This means, that inverses have to
be established for any grade by direct computation or other algebraic means.
\myeenv

We would like to remark at this point, that it is this failure of complete 
multiplicativity which was the origin of the present investigation. In 
quantum field theory one needs to consider renormalization. There, certain 
integrals need to be regularized by a subtraction process called
renormalization which includes compensating singular integrals, the counter
terms. There are strong hints, that the failure of total multiplicativity here
and its re-establishment using `counter terms' has a deep link to this 
renormalization process. The work where this might appear to be seen at least
with hindsight is the work of Brouder and Schmitt
\cite{brouder:schmitt:2002a}, see also section \ref{sec:Renormalization}
and \cite{brouder:fauser:frabetti:oeckl:2002a}. To remedy the situation 
we have two options (at least). 

\noindent
{\bf Option A:} We know the following, using the Oziewicz crossing
\cite{oziewicz:1997a,oziewicz:2001b,fauser:oziewicz:2001a,fauser:2002c}. 
Analogous to theorem \ref{addOziewicz} we find:
\mybenv{Theorem} The antipodal convolution is a \textbf{Hopf convolution}
and hence a \textbf{Hopf gebra} with respect to the following crossing 
\begin{align}
(\cdot\ot \cdot)(\Id \ot \DM \ot \Id)(\sm\ot \Id\ot \sm)
(\Id\ot \cdot \ot \Id)(\DM\ot \DM)
\end{align}
\myeenv
\noindent
{\bf Proof:} The failure for being Hopf has been demonstrated above. The
existence of this crossing is guaranteed, due to biassociativity and the 
existence of the antipode.\qed

The Dirichlet convolution ring thus forms a 
Hopf gebra only. It needs further checking if this Hopf gebra could still 
be braided to form a braided Hopf \textit{al}gebra. It would then be possible
to develop this crossing as the ordinary switch plus additional correction
terms $\tau=\textsf{sw}+O(1)$. In \cite{fauser:oziewicz:2001a} we had to 
learn that this route is rather tedious.

\noindent
{\bf Option B:} We can use a slightly altered version of the renormalized
multiplication coproduct, which adjusts the multiplicities which are 
responsible for the failure of being multiplicative. Such an `unrenormalized' 
coproduct would emerge from a different type of duality, and we are lead to
a new duality map beside the Kronecker duality used above. A candidate for such
a pairing turns out to be the 
\textbf{unrenormalized Kronecker pairing}\footnote{
This name is chosen using wishful thinking and awaited applications in pQFT, 
see also \cite{brouder:schmitt:2002a} and section \ref{sec:Renormalization}.
It depends however on the viewpoint which of the coproducts should be 
addressed as `renormalized'. Our naming scheme reflects the usage of the term
renormalized in physics. Further more, the Dirichlet convolution is the
mathematical interesting structure and not the unrenormalized one, hence our
awkward development starting using the unrenormalized coproducts.
}
\begin{align}
(n\mid m) 
&= R(n,m)=\delta_{n,m}\prod_i r_i! = (\prod_i r_i!)\,\langle n\mid m \rangle
\end{align}
where $n=\prod_i p_i^{r_i}$. This pairing dualizes multiplication by
employing $R$ in eqn. (\ref{renKroneckerDual}) in such a way that the pair 
($\cdot,\DR$), see definition \ref{DR} below, fulfils the \textbf{Hopf algebra} 
axioms. The precise definition will be given in theorem (\ref{renKroneckerP}).
Furthermore, our renormalized multiplication coproduct becomes a 
relative of the renormalization coproduct of \cite{brouder:schmitt:2002a}. We 
will discuss this in more detail below.

\subsection{The unrenormalized coproduct}

We noticed above, that the coproduct obtained from dualizing multiplication
is only an homomorphism if we consider relatively prime integers. Hence this
property does not hold for the monoid $\ZP$ as a whole. There is a standard
method to circumvent this problem, and to define an actual Hopf algebra in
such a way that the new coproduct is a homomorphism on $\ZP$. This will turn
out to be the unrenormalized coproduct.

\mybenv{Definition}
The monoid $\ZP$ can be graded by the number of primes of every $n\in \ZP$,
where multiplicities are counted. Let $n=\prod_i p_i^{r_i}$, $\nu=\sum_i r_i$,
\begin{align}
\ZP &= 1+ \ZP^1 + \ZP^2 + \ldots \ZP^{\nu} +\ldots \nn
\ZP^{\nu} &= \langle \prod_j p_j^{r_j}\rangle 
             \quad\text{where~} \sum r_i=\nu \nn
\ZP^{\nu} \cdot \ZP^{\nu^\prime} &= \ZP^{\nu+\nu^\prime}
\end{align}
\myeenv

The grade one elements, i.e. the primitive elements, are the infinitely many
primes $p_i$. We can use now the Kronecker duality to dualize the 
multiplication of a single prime $p$ and obtain as above 
$\DM(p)=p\ot 1 + 1\ot p$. A coproduct which fulfils the 
homomorphism property can now defined by \textit{recursion} on the grade
forcing the homomorphism property.

\mybenv{Definition}\label{DR}
The \textbf{unrenormalized coproduct of multiplication} $\DR$ is defined 
recursively on the graded monoid $\ZP$ as
\begin{align}
\DR(\etam)&=\etam\ot\etam &&\text{on~}\ZP^0\nn
\DR(p) &= p\ot 1 + 1\ot p &&\text{on~}\ZP^1\nn
\DR(m\cdot n) &= \DR(n)\DR(m) &&\text{otherwise,}
\end{align}
where $\ZP^0=1$ and $\ZP^1=\{p_i\}$ and $p_i$ are prime numbers.
\myeenv

\mybenv{Corollary}
The explicit form of the unrenormalized coproduct $\DR$ on $n=\prod^k_i
p_i^{r_i}$ is given as
\begin{align}\label{cop}
\DR(n)
&=
\prod_i \DR(p_i^{r_i}) \nn
\DR(p_i^r) 
&= 
\sum_{s=0}^r {r \choose s} p_i^{r-s} \ot p_i^s \nn
\DR(n)
&=
\sum_{s_1}^{r_1}\ldots \sum_{s_k}^{r_k}
{r_1\choose s_1}\ldots{r_k\choose s_k}
p_{i_1}^{s_1}\ldots p_{i_k}^{s_k}\ot
p_{i_1}^{r_1-s_1}\ldots p_{i_k}^{r_k-s_k}
\end{align}
The $(r-1)$-iterated coproduct of a single prime power $p^r$ is given as
\begin{align}\label{iterDR}
\DR^{(r-1)}(p^r) &= \nn
\sum_{s_1=0}^r \sum_{s_2=0}^{s_1} &\ldots\sum_{s_{r-1}=0}^{s_{r-2}}
{r \choose s_1}{s_1 \choose s_2}\ldots{s_{r-1} \choose 0}
p^{r-s_1}\ot p^{s_1-s_2} \ot \ldots \ot p^{s_{r-1}}
\end{align}
\myeenv
\noindent
{\bf Proof:} We know by definition that the coproduct is a homomorphism, hence
we need to consider prime powers only. From 
\begin{align}
\DR(p^r) 
&= (\DR(p))^r = (p \ot 1 + 1 \ot p)^r \nn
&= \sum_{s=0}^r {r \choose s} p^s \ot p^{r-s}
\end{align}
we get the first part of the assertion. The second assertion follows by
iteration of the coproduct.\qed

Note that the appearance of the \textit{binomial coefficients} makes up the 
difference between the renormalized coproduct of multiplication and the 
unrenormalized coproduct of multiplication.
\begin{align}
\DM(p^2) 
&= \sum_{d\mid 2} p^d \ot p^{\frac{2}{d}}
 = p^2 \ot 1 + p\ot p + 1 \ot p^2 \quad\text{since~} d\in \{1,p,p^2\} \nn
\DR(p^2)
&= \sum_{s=0}^2 {2 \choose s} p^s \ot p^{2-s} 
 = p^2 \ot 1 + 2\,p\ot p + 1 \ot p^2  
\end{align}
Hence this change actually involves a duality and employs the divided power
structure. It should be noted that the iterated coproduct of eqn. 
(\ref{iterDR}) has an analogous structure to a path-ordered product,
see \cite{cartier:2000a} for a discussion.

We are now able to define a new scalar product based on the Laplace pairing,
see \cite{rota:stein:1994a,brouder:fauser:frabetti:oeckl:2002a,fauser:2002c},
induced by the renormalized coproduct. The Laplace pairing is obtained by 
an expansion formula analogy to the expansion of a determinant. One uses the
fact that the product can be dualized into the coproduct an vice versa. In 
our case we demand that
\begin{align}
(\DR(p)\mid r\ot s) &= (p\mid r\cdot s) \nn
(p\ot q\mid \DR(s)) &= (p\cdot q\mid s) 
\end{align}
Of course, this cannot be the Kronecker pairing since we know that the pair
($\cdot,\DR$) is not Kronecker dual.
\mybenv{Theorem}\label{renKroneckerP}
The \textbf{unrenormalized Kronecker pairing} $(.\mid .) : \ZP \times \ZP
\rightarrow \ZP$ is given by Laplace expansion of $n=\prod_i p_i^{r_i}$ and
$m=\prod_j p_j^{s_j}$ as
\begin{align}
(n\mid m) 
&= \prod_i \delta_{r_i,s_i} r_i!
\end{align}
\myeenv
\noindent
{\bf Proof:} We need to show that the pairing can be defined recursively by 
Laplace expansion of either $n$ or $m$. To do so we assume $n\ge m$ ($n<m$
follows by symmetry) and denote $q_i = p_i^{s_i}$. Then we compute 
\begin{align}
(n\mid m) 
&= (n \mid q_1\cdot \ldots \cdot q_k ) \nn
&= (\DR^{(k-1)}(n) \mid q_1 \ot \ldots \ot q_k) \nn
&= (n_{(1)}\ot \ldots \ot n_{(k)}\mid q_1 \ot \ldots \ot q_k) \nn
&= (n_{(1)}\mid q_1)^r(n_{(2)}\mid q_2)^r\ldots(n_{(k)}\mid q_k) 
\end{align}
Furthermore we have
\begin{align}
(p_i^{r_i}\mid p_j^{s_j}) 
&=(\DR^{(s_j-1)}(p_i^{r_i}) \mid p_j\ot \ldots \ot p_j) \nn
&=\sum_{s_1=0}^{r_i}\sum_{s_2=0}^{s_1}\ldots \sum_{s_{r_i-1}}^{s_{r_i-2}}
{r\choose s_1}{s_1\choose s_2}\ldots{s_{r_i-2}\choose s_{r_i-1}} \nn
&~~~~~~(p_i^{r_i-s_1} \mid p_j)(p_i^{s_1-s_2}\mid p_j) \ldots 
(p_i^{s_{r_i-2}-s_{r_i-1}}\mid p_j) \nn
&= \delta_{i,j}\delta_{r_i,s_i} r_i \cdot (r_i-1) \cdot \ldots \cdot 1
\end{align}
since only terms of the form $(p_i\mid p_j)$ for $i=j$ survive.\qed

The counit $\epm$ of $\DM$ is still the counit of $\DR$, the product remains 
unchanged and hence also the convolutive unit $\up$ is unaltered. We can 
compute the unrenormalized antipode for the convolution $(\cdot,\DR)$. 
\mybenv{Theorem}\label{sr}
The unrenormalized antipode $\sr$ of the unrenormalized multiplicative
convolution on an element $n=\prod_i p_i^{r_i}$ of grade $\nu=\sum_i r_i$
is given as $\sr(1)=1$ and for $n\ge 1$
\begin{align}
\sr(n) &= (-1)^\nu n
\end{align}
\myeenv
\noindent
{\bf Proof:} We compute recursively the defining equation of the antipode
$(\sr\conv \Id)(n) = \etam\circ\ep(n) = \delta_{n1}\etam$
\begin{align}
n=1) && \sr(1)1=1 &&& \sr(1)=+1 \nn
n=2) && \sr(2)+\sr(1)2=0 &&& \sr(2)=-2 \nn
n=3) && \sr(3)+\sr(1)3=0 &&& \sr(3)=-3 \nn
n=4) && \sr(4)+2\sr(2)2+\sr(1)4=0 &&& \sr(4)=+4 \nn
n=5) && \sr(5)+\sr(1)5=0 &&& \sr(5)=-5 \nn
n=6) && \sr(6)+\sr(3)2+\sr(2)3+\sr(1)6=0 &&& \sr(6)=+6 \nn
\vdots && \nn
n=8) && \sr(8)+3\sr(4)2+3\sr(2)4+\sr(1)8=0 &&& \sr(8)=-8 \nn
\vdots && 
\end{align} 
The cases $n=4$ and $n=8$ show clearly, that the non square-free numbers now
also get a sign grading, and no longer disappear. From this observation and
the homomorphism property of the coproduct the conclusion can be drawn.\qed

\mybenv{Corollary}
The $5$-tuple $H=(\ZP,\cdot,\etam,\epm,\sr)$ is a Hopf algebra.
\myeenv
This is true by construction.

\mybenv{Corollary}
The unrenormalized antipode $\sr$ is involutive $\sr^2=\Id$ as a linear 
operator.
\myeenv
This has topological consequences, see \cite{kuperberg:1991a}.

\subsection{\label{sec:3-4}Branching operators and division}

A branching operator for the multiplicative convolution is now defined along
the same lines as described above in definition \ref{boa}, consult also
\cite{brouder:fauser:frabetti:oeckl:2002a,fauser:jarvis:2003a}. 
\mybenv{Definition}\label{bom}
A \textbf{branching operator} for the multiplicative convolution is given
by a 1-cochain $\phi$ and the coproduct as
\begin{align}
/\Phi 
&= \cdot(\phi \ot\Id)\DM 
 = \cdot(\Id \ot \phi)\DM \nn
/\Phi(n) 
&=\phi(n_{[1]})n_{[2]} = \phi(n_{[2]})n_{[1]} 
\end{align} 
\myeenv
We have of course now two duality maps available to define special 1-cochains
parameterized by elements $b$ in $\ZP$. These are the renormalized Kronecker 
duality $K$ and the unrenormalized Kronecker duality $R$. Hence we have two 
possibilities $\phi_b(n)=\langle b\mid n\rangle$ and $\phi^r_b(n)= (b\mid n)$ 
to obtain a branching. The first gives us the

\mybenv{Theorem}\label{bomdiv}
The branching operator $/\Phi_b$ w.r.t. the 1-cochain $\phi_b$ acts
as division by $b$ if the argument is divisible by $b$
and as projection to $0$ otherwise.
\myeenv
\noindent
{\bf Proof:} We compute the branching operator as
\begin{align}
/\Phi_b(n) 
&= (\phi_b \ot \Id)\DM(n) \nn
&= \phi_b(n_{[1]}) n_{[2]} = \sum_{d\mid n} \phi_b(d) \frac{n}{d} \nn
&= \sum \delta_{b,d} \frac{n}{d} = \left\{
\begin{array}{cl}
{\displaystyle\frac{n}{b}} & \text{if $b\mid n$} \\
0   & \text{otherwise}
\end{array}\right.
\end{align}
showing the desired feature.\qed

We might wonder, if the branching operators of primitive elements fulfil a
Leibniz type rule. Indeed we can derive the following
\mybenv{Corollary}\label{corLeibniz}
Let $\epm=\phi_1$ be the trivial 1-cochain and $\phi_p(n)=\delta_{p,n}$ a
1-cochain based on a primitive element $p$ (prime number) with branching
operator $/\Phi_p$. Let $n,m$ be relatively prime; then one has the
Leibniz rule
\begin{align}
/\Phi_p(n\cdot m) &= 
/\Phi_{p[1]}(n_{[1]})/\Phi_{p[2]}(m_{[1]})\, n_{[2]}\cdot m_{[2]} \nn
&= n\cdot /\Phi_p(m) + /\Phi_p(n)\cdot m
\end{align}
\myeenv
This result is however somehow void. Since $n,m$ are relatively prime, a
factor containing $p$ occurs either in $n$ or in $m$ or in neither term,
and hence one of the two terms or both vanishes, but in general not the
other one. If $n,m$ are not relatively prime the result does not hold,
since we needed the homomorphism axiom to come up with it. However,
division is not expected to be a derivation at all. 

\subsection{\label{sec:3-5}Branching operators and derivation}

The second option is employed in a disguised version. Having the unrenormalized 
coproduct at our disposal, we are now in the position to introduce a different
branching. We can compose the unrenormalized coproduct and the renormalized
Kronecker evaluation map to define a new type of action of elements on the 
monoid $\ZP$. This can be seen in two different ways, either by changing the 
identification between the monoid and its dual, or by keeping this canonical 
isomorphism and changing the coproduct.
\begin{align}
R(V,V) &\cong \eval(R(V)\otimes V) \cong \eval(V^{\#} \otimes V) \nn
K(V,V) &\cong \eval(K(V)\otimes V) \cong \eval(V^* \otimes V)
\end{align}
where $R: V \rightarrow V^{\#}$ is the duality induced by the
unrenormalized pairing and $K$ is the same map induced by the 
renormalized Kronecker pairing, recall definitions \ref{renKroneckerDual} 
and \ref{KroneckerDual}. 

To start, we choose the second possibility keeping evaluation and 
Kronecker pairing straight. Using the renormalized coproduct we get 
another interesting map, a contraction map $\JJ{R}$. Contractions are
related to branching operators as
\begin{align}
\JJ{\relax} : V \otimes V \rightarrow V 
&\Leftrightarrow
/\Phi_V : V \rightarrow V \nn
n \JJ{\relax} m \cong i_{n}(m) \cong K(n)(m)
&\rightarrow 
K(b) \cong /\Phi_b
\end{align}
explicitly
\begin{align}
n \JJ{R} m 
&= (\eval\ot \Id)(K \ot \DR)(n\ot m) = /Phi^R_n(m)
\end{align}
This map has the properties of a \textit{derivation} as we may calculate
for $p,q$ different primes
\begin{align}
p \JJ{R} (p^2q) 
&= (\eval \ot \Id)
   \big( (p^*,p^2q,1)+2(p^*,p,pq)+2(p^*,pq,p) \nn
&~\hskip 1truecm +(p^*,q,p^2)+(p^*,p^2,q)+(p^*,1,p^2q) \big) \nn
&= 2 pq
\end{align}
We have used $K(p)=p^*$ as the linear form attached to $p$. We 
consider once more $\ZP$ as a graded space, the grading induced by the number
of primes constituting a number. Note that only grade one elements act 
as derivations obeying a Leibniz rule while higher grade elements act due to
the Hopf algebra structure, see \cite{fauser:2002d}. Since we use the
evaluation map and the renormalized coproduct obtained from a different
pairing, this Hopf algebra structure is \textit{no longer selfdual}.
This may be summarized as

\mybenv{Theorem}
The branching operator $/\Phi^{R}_{p_i}$ w.r.t. the $1$-cochain 
$\phi_{p_i}$, $p_{i}$ a primitive element (prime number) acts as a 
derivation by $p_i$ if the argument contains a factor $p_i$ (a non 
constant function of $p_i$) and as projection to $0$ otherwise.
\myeenv
\noindent
{\bf Proof:} We compute, using $n=\prod p_i^{r_i}$ and the unrenormalized
coproduct, the branching operator as
\begin{align}
/\Phi^{R}_{p_i}(n) 
&= (\phi_{p_i} \ot \Id)\DR(n) \nn
&= \phi_{p_i}(n_{[1]}) n_{[2]} 
 = \sum {r_1 \choose s_1}\ldots{r_k \choose s_k}
   \phi_{p_i}(\prod p_j^{s_j}) \prod p_j^{r_j-s_j} \nn
&= \sum r_i \delta_{p_i,p_j} \frac{n}{p_i} = \left\{
\begin{array}{cl}
r_i \frac{n}{p_i} & \text{if $p_i \mid n$} \\
0   & \text{otherwise}
\end{array}\right.
\end{align}
showing the desired feature.\qed

\mybenv{Corollary}
A branching by a primitive element fulfils the Leibniz rule.
\myeenv
\noindent
{\bf Proof:} Since the homomorphism axiom holds, the proof is standard, see
proof given for the division, corollary \ref{corLeibniz}.\qed

Hence a branching operator can be identified somehow with a derivation 
$/\Phi^R_{p_i}(n)= \partial_{p_i}\,n$. This can bee seen from
\begin{align}
\partial_{p_i}\,n &= \partial_{p_i}\prod p_j^{r_j} \nn
&= r_i \prod p_j^{r^\prime_j},\quad \text{where~} r_j^\prime=r_j,~i\not=j,
\text{~and~} r_i^\prime = r_i-1 \text{~or~} 0 \text{~if~} r_i=0.
\end{align}
The proof of the preceding theorem shows that the alternative way to 
define the branching operators using the renormalized pairing and the 
unrenormalized coproduct property gives an equivalent result
\begin{align}
/\Phi^{R}_p(n) 
&= (\phi_p\ot \Id)\DR(n) \nn
&= (\phi^{R}_p\ot \Id)\DM(n)
\end{align}
since $\phi^{R}_{p_i}(d) = ( p_i\mid d) = r_i \delta_{p_i,d}$ and
$\phi_{p_i}= \langle p_i\mid d\rangle = \delta_ {p_i,d}$. Hence we can
identify $\phi^R_b = b^{\#}$ and $\phi_b=b^*$. This shows, that we have 
two alternatives to incorporate the unrenormalized structure, via the 
coproduct or via the pairing. The hereby used linear form is the 
exponential of the grade one linear form.

\subsection{Algebraic identification of the duals}

We introduced duality maps $K$ and $R$ in definitions \ref{KroneckerDual},
\ref{renKroneckerDual}. These maps induce an isomorphism between 
the space $V$ and $V^*$ or $V^{\#}$. This can be used to dualize the
coproduct structures on the mutual dual spaces. 
\mybenv{Theorem}
Given the unrenormalized Kronecker coalgebra of multiplication. Under 
the dualities $K$ and $R$ the induced multiplications on $V^*$ and 
$V^{\#}$ the renormalized duality $K$ yields the unrenormalized Kronecker 
multiplication, which is again ordinary multiplication, and the 
unrenormalized dualization $R$ implies a divided power multiplication.
\myeenv 
\noindent

{\bf Proof:} We denote the image of an element $a$ under the map $K$ 
as $K(a)=a^*$ and the image of $a$ under the map $R$ as $R(a)=a^{\#}$. 
We employ the universal action of the dual, the evaluation $\eval$, 
and compute  
\begin{align}
\eval(n^*\cdot m^* \ot k) 
&= \langle n\cdot m \mid k\rangle = \delta_{n\cdot m,k}
\end{align}
and using $k=\prod_i p_i^{r_i}$ we get
\begin{align}
\eval(n^{\#}\cdot m^{\#} \ot k) 
&= (n\cdot m \mid k) = \prod_i r_i !\delta_{n\cdot m,k}
\end{align}
Using $n=\prod_i p_i^{r_i}$ and $m=\prod_j p_j^{s_j}$ we might write 
(formally) $n^{\#}= (\prod_i r_i!)^{-1} n$ and compute
\begin{align}
n^{\#} \cdot m^{\#} 
&= (\prod_i r_i!)^{-1} (\prod_i s_i!)^{-1} n \cdot m \nn
&= \frac{(\prod_i (r_i+s_i)!)}{\prod_i r_i!\prod_j s_j!}
 (\prod_i (r_i+s_i)!)^{-1} n\cdot m\nn
&=\frac{(\prod_i (r_i+s_i)!)}{\prod_i r_i!\prod_j s_j!}
 \,(n\cdot m)^{\#}
\end{align}
showing that the duals obtained by the renormalized Kronecker pairing
forms a divided powers algebra, compare section \ref{sec:Occupation}
and eqn. (\ref{diag1}). Examples of such multiplications are
$2^{\#}\cdot 4^{\#} = 3\cdot 8^{\#}$ and $4^{\#}/2^{\#} = 1/2\cdot 2^{\#}$,
which we will have occasion to need in section \ref{sec:SymFct}.
\qed

This is in accord with eqn. (\ref{SymDivp}). It should be noted, that 
the divided powers algebra $\divp(P)$, over the space $P$ spanned by 
primitive elements $p_i$ cannot be generated multiplicatively in 
characteristic $\not=0$. E.g. $1^{\#}\cdot 1^{\#} = 2\cdot 2^{\#}$ 
where $2$ is not necessarily a unit. To remedy this one can introduce
a Rota-Baxter operator $R$ of weight $0$ as employed in section 
\ref{sec:Occupation}, which establishes $R(n^{\#})=(n+1)^{\#}$.

\subsection{\label{sec:DGF}Dirichlet series, Dirichlet $L$-series, 
Dirichlet generating functions} 

In several places we have colloquially introduced `functions' $f(s)$
in connection with formal series $f_1, f_2, f_3, \ldots$ without further
details. For example, the M\"obius function $\mu(s)$ is of this type. 
These notions shall be made more precise in this subsection.
\mybenv{Definition}
A \textbf{Dirichlet generating function} is a formal power 
series constituting an arithmetic function in the following form 
\begin{align}
f(s) 
&= \sum_{n\ge 1} \frac{f_n}{n^{s}}
\end{align}
where $f_n,s=\sigma+i\tau\in\openC$ are complex numbers. 
\myeenv
\mybenv{Examples}
\begin{align}
\begin{array}{ccc}
f(s) & f_n & \text{sequence} \\
\hline\hline
\zeta(s) & 1 & 1,1,1,1,1,1,1,1,1,1,\ldots \\
\zeta(s)^2 & d(n) & 1,2,2,3,2,4,2,4,\ldots \\
\frac{1}{2-\zeta(s)} & H(n) & 1,1,1,2,1,3,4,2,3,1,8,\ldots \\
\lambda(s) & \frac{1}{2}(1-(-1)^n) & 1,0,1,0,1,0,1,0,1,0,1,0,\ldots\\
\mu(s) & \mu_n & 1,-1,-1,0,-1,1,-1,0,0,1,\ldots \\
\hline
\end{array}
\end{align}
Where $\zeta(s)$ is the Riemann zeta-function, $\zeta(s)^2$ is the divisor
function, $d_n=d(n)$ is the number of divisors of $n$, $H(n)$ is the number of
ordered factorizations of $n$, $\lambda(s)$ is the Dirichlet lambda-function
and $\mu(s)$ is the M\"obius function. Note, that these functions may or may
not be (complete) multiplicative.
\myeenv
This definition can be generalized using number theoretic characters
$\chi_k(n)$, which are complete multiplicative $k$-periodic functions,
$\chi_k(n+k)=\chi_k(n)$.
\mybenv{Definition}
A \textbf{Dirichlet $L$-series} is a Dirichlet series of the form
\begin{align}
L_k(s,k) &= \sum_{n\ge 1}\chi_k(n) n^{-s}
\end{align}
where the number theoretic character $\chi_k$ is an integer function with
period $k$.
\myeenv
There is a close connection here to modular forms etc. on which we will 
not dwell, as we in general refrain here from making number theoretic 
statements. We are now able to use some basic facts to obtains a more
concise form of the formula for the renormalized antipode.

\mybenv{Theorem}
The M\"obius function and the Riemann zeta-function are mutually convolutive
inverse Dirichlet series.
\begin{align}
(\mu\conv\zeta)(s) 
&= \sum_{n\ge1 } \sum_{d\mid n} \mu_d \zeta_{n/d} n^{-s} \\
&= \sum_{n\ge1 } \delta_{n1} n^{-s} = 1
\end{align}
\myeenv
\noindent
{\bf Proof:} Its well known that $\sum_{d\mid n} \mu_d=\delta_{n1}$. Another
interesting proof uses the \textbf{Euler product} of the Dirichlet series.
This is a product over all primes $p_i$
\begin{align}
\frac{1}{\zeta(s)} 
&=\prod_{i\ge 1} (1-\frac{1}{p_i^{s}}) \nn
&=(1-\frac{1}{p_1^{s}})(1-\frac{1}{p_2^{s}})(1-\frac{1}{p_3^{s}})\ldots \nn
&=1-(\frac{1}{p_1^{s}}+\frac{1}{p_2^{s}}+\frac{1}{p_3^{s}}+\ldots) \nn
&~+(\frac{1}{p_1^sp_2^s}+\frac{1}{p_1^sp_3^s}+\ldots+
    \frac{1}{p_2^sp_3^s}+\frac{1}{p_2^sp_4^s}+\ldots + )
  -\ldots \nn
&=1-\sum_{i\ge1} \frac{1}{p_i^s} 
   +\sum_{j>i\ge 1} \frac{1}{p_i^sp_j^s}
   -\sum_{k>j>i\ge 1} \frac{1}{p_i^sp_j^sp_k^s}
   + \ldots \nn
&= \sum_{n\ge 1} \frac{\mu_n}{n^s}
\end{align}
Since we do only formal algebra here, we need not to bother about convergence
and the claim is proved.\qed 

The reader should note the quite close relation ship between series of
symmetric functions and the Euler products. In fact, one has $E =
\prod_{i\ge 1} (1+x_i)$ as the generating function of the elementary
symmetric functions. The product form yields the series 
\begin{align}
\prod_{i\ge1} (1+x_i) 
&=
\sum_{n\ge 0} e_n(x_1,x_2,\ldots) 
\end{align}
and Dirichlet series emerge this way as a specialization of the elementary 
symmetric functions on the values $x_i=-p_i^{-s}$ of complex prime powers. 
Since one knows, that such specializations assign the energy
eigenvalues of a Hamiltonian system $x_i=e^{E_i t}$, this is in accord with
recent attempts to find (quantum) Hamiltonian systems having a `Riemann
operator' which generates the zeros of the Riemann zeta function. However, 
our approach is somehow inverse to that.

\mybenv{Theorem}
The antipode of the Kronecker multiplication coproduct convolution is given 
by the shifted M\"obius Dirichlet series $\mu(s-1)$
\myeenv
\noindent
{\bf Proof:} We derived in theorem \ref{antipm} that the antipode had as 
coefficients the Ha\-da\-mard product of the identity series $\Id_n=n$ and
that of the M\"obius series $\mu_n$. This amounts to saying
\begin{align}
\label{dirichletantip}
\sm(s) &= (\Id.\mu)(s) = \sum_{n\ge1} \frac{n\cdot\mu_n}{n^s} 
        = \sum_{n\ge1} \frac{\mu_n}{n^{s-1}}\nn
       &= \mu(s-1)
\end{align}
showing the claim.\qed

It should be noted, that Dirichlet series converge for sufficiently large real
part of the complex parameter $s$, hence the antipode using $s-1$ is more 
singular that the M\"obius function itself, converging only for $\Re(s)>2$. 
Also note that the identification of series as functions is not unique. 
Before closing this subsection we should mention  that the M\"obius series, 
for example, can be expanded as a Lambert series too. 
\begin{align}
\Lambert(a_n,x) &= \sum_{n\ge 1} a_n\frac{x^n}{1-x^n} \nn
\Lambert(\mu_n,x) &= \sum_{n\ge 1} \mu_n  \frac{x^n}{1-x^n} = x
\end{align}
This gives an additional interesting functional relation attached to the
M\"obius series and implies other coalgebraic structures, which we cannot 
investigate here. Another potential generalization of Dirichlet series is 
the Hurwitz-Riemann zeta-function, given as
\begin{align}
\zeta_k(s)&= \sum_{n\ge 1} \frac{1}{(n+k)^s}
\end{align}
Note, that the Dirichlet lambda-function can be written as
\begin{align}
\lambda(s) 
&= \sum_{n\ge1} \frac{1}{(2n+1)^s} 
 = (1-2^{-s}) \zeta(s). 
\end{align}

\subsection{Polylogarithms}

As a short but interesting aside, we consider another Dirichlet-like 
series of interest in physics, the polylogarithms. 
\begin{align}
\Li_s(z) &= \sum_{n\ge1} \frac{z^n}{n^s}
\end{align}
For $z=1$ one stays with the Riemann zeta-function. In the sense of
combinatorics, this function should be looked at as a twofold generating
series, where one part is of ordinary polynomials series type and the
other is of Dirichlet type. In this sense we should study series of the
form of an ordinary series-Dirichlet generating function (odg)
\begin{align}
\odg(a_{n,m},z,s) &= \sum_{n\ge1,m\ge0} a_{n,m} \frac{z^m}{n^s}
\end{align}
which is the product of two such series. The polylogarithms are then
those series for which the $a_{n,m}=\delta_{n,m}$. The scheme presented
here encodes mappings from the additive into the multiplicative
realm and vice verse, hence exponential and logarithm maps respectively. 

Polylogarithms occur as complete integrals over Fermi-Dirac and
Bose-Einstein distributions while evaluating Feynman diagrams.
The generalization to many variables, i.e. to Euler-Zagier sums 
will not be considered, but emerges as a natural generalization 
of the present structure to (co)monoids of the form $\ZP^\ot$.

\section{Relation between addition and multiplication}
\setcounter{equation}{0}\setcounter{mycnt}{0}

\subsection{The renormalized case}

From arithmetic it is clear that addition and multiplication have an 
intimate relation. This can be made even clearer using categorial methods
\cite{lawvere:rosebrugh:2003a}. First we recognize, that multiplication
is the repeated application of addition. Define an operator
$\add_n : \ZP\rightarrow \ZP$, $\add_n(m)=n+m$ then we find
\begin{align}
\cdot(n,m) &= \add_n^m(0)
\end{align}
This suggests, that the coproduct of addition $\DP$ and the coproduct of
multiplication $\DM$ have to be related by similar arguments in a dual 
fashion. Indeed we find very easily the following relation
\mybenv{Theorem}\label{DP2DM}
The renormalized coproduct of multiplication $\DM$ is related to the 
renormalized coproduct of addition $\DP$ by exponentiation
\begin{align}
n
&= \prod_i p_i^{r_i} \nn
\DM(n) 
&= \delta^\DM (n) 
 = \sum_{r_i^\prime+r_i^{\pprime}=r_i,\forall i} \prod_i p_i^{r_i^\prime}
\ot \prod_i p_i^{r_i^\pprime}
\end{align}
where $\delta$ is the group like coproduct $\delta(n)=n\ot n$ on primes and
the $r_i^\prime$, $r_i^\pprime$ run in $\ZP$ (including zero). 
\myeenv
\noindent
{\bf Proof:} The theorem is clear for $1=p_i^0$ and for a single prime 
$\DM(p)= p\ot \etam + \etam\ot p = \sum_{r+s=1} p^r\ot p^s$. Since the
coproduct is not in general multiplicative, we need to check the relation for
all $p^r$ separately.
\begin{align}
\DM(p^r) 
&= \sum_{d\mid p^r} d \ot \frac{n}{d} \nn
&= \sum_{d=p^s,s=0}^{r} p^s \ot p^{r-s} \nn
&= \delta(p)^{\DM(r)} = (p\ot p)^{\DM(r)} \nn
&= \sum_{s+t=r} p^s \ot p^t
\end{align}
Now, all integers have a unique prime number factorization, so we have
$n=\prod_i p_i^{r_i}$. Furthermore we know that the coproduct of multiplication
is a homomorphism of multiplication for \textit{relatively prime} integers
$n,m$. Hence we find the important formula
\begin{align}
\DM(n) 
&=\DM(\prod_i p_i^{r_i}) = \prod_i \DM(p_i^{r_i}) \nn
&= \prod_i \delta(p_i)^{\DP(r_i)} \nn
&= \sum_{r_i^\prime+r_i^\pprime=r_i,\forall i}
  \prod_i p_i^{r_i^\prime} \ot \prod p_i^{r_i^\pprime}
\end{align}
which proves the theorem.\qed

As a mnemonic we may state that $\DM=\delta^\DP$, the assumed exponential
relationship we were after. There would be much need to say more about
distributivity and its failure, but we postpone this for further
investigations. 

\subsection{The unrenormalized case}

The unrenormalized case is best treated reversing the argumentation. We 
already have a formulation of the unrenormalized coproduct in terms of 
an additive expression in the exponents of the two factors, see eqn. 
(\ref{cop}). We make the
\mybenv{Definition}
The unrenormalized coproduct of addition $\DPR$ is given as
\begin{align}
\DPR(n) &= \sum_{0\le r\le n} {n \choose r} r\ds (n-r)
\end{align}
\myeenv
which allows us to state the 
\mybenv{Theorem}\label{DRP}
The unrenormalized coproduct of multiplication $\DR$ is related to the
unrenormalized coproduct of addition $DPR$ by exponentiation
\begin{align}
n
&= \prod_i p_i^{r_i} \nn
\DR(n) 
&= \delta^\DPR (n) 
 = \sum_{r_i^\prime+r_i^{\pprime}=r_i,\forall i} 
{r_i^\prime+r_i^{\pprime} \choose r_i^\prime}
\prod_i p_i^{r_i^\prime}
\ot \prod_i p_i^{r_i^\pprime}
\end{align}
where $\delta$ is the group like coproduct $\delta(n)=n\ot n$ on primes and
the $r_i^\prime$, $r_i^\pprime$ run in $\ZP$ (including zero). 
\myeenv
In fact this is at the same time a definition of the unrenormalized 
coproduct of addition. The difference between the unrenormalized and the 
renormalized coproducts is given by the binomial prefactors. The way 
in which way linearity is to be handled is a somewhat delicate point.
We could as prefactor adopt the exponent of the rhs. in
\begin{align}
{r_i^\prime+r_i^{\pprime} \choose r_i^\prime} 
&=
p_i^{\ln_{p_i}({r_i^\prime+r_i^{\pprime} \choose r_i^\prime})}
\end{align}
where the $\ln$ is taken with respect to the base $p_i$. From the point of
view of a characteristic free development, this looks unnatural. The proper
definition of the unrenormalized coproduct of addition therefore requires
further work, with a relation to the von Mangtoldt function $\Lambda(s)$ 
expected.

\subsection{\label{sec:Witt}Lambda ring structure, Witt vectors and the 
Witt functor}

This section is somewhat more abstract and not strictly needed for the
applications in section \ref{sec:Applications}. In a characteristic
free development one needs these considerations.

Thus far, we have implicitly used more advanced constructions then just addition
and multiplication. In this subsection we make some of this mathematics explicite.
Our aim is to present development in other areas of mathematics and physics, and
induced there is an underlying categorial background which guarantees the
universality of our results. For example, it is well known that the binomial
coefficients are related to a lambda ring structure on the ring $\openZ$ of all
integers. Here we elaborate on this in order to set the scene for generalizations
of applications presented in the next section to finite characteristics. We begin
by recalling a few facts from Knutson's book about lambda rings
\cite{knutson:1973a}. Given a ring $\bfR$, we associate to it in a functorial 
way a new ring $1+\bfR\PS{t}^+$, where $\bfR\PS{t}^+$ is the augmentation ideal, 
i.e. the kernel of the counit, that is to say power series in $t$ with a zero 
constant term. This ring is defined by the pair $(\bfR,\lambda_t)$ called a 
lambda ring. The lambda operations fulfil for all $x,y\in\bfR$
\begin{align}
\lambda^0(x) &= 1 \nn
\lambda^1(x) &= x \nn
\lambda^n(x+y) &= \sum_{r=0}^n \lambda^r(x)\lambda^{n-r}(y) \nn
\lambda^n(xy) &= 
\textsf{P}_n(\lambda^1(x),\lambda^2(x),\ldots,\lambda^n(x),
\lambda^1(y),\lambda^2(y),\ldots,\lambda^n(y)) \nn
\lambda^n(\lambda^m(x)) &=
\textsf{P}_{n\cdot m}(\lambda^1(x),\lambda^2(x),\ldots,
\lambda^{n\cdot m}(x)) \nn
\lambda_t(1) &= 1+t,\qquad \lambda_t(x) =\sum \lambda^n(x)t^n
\end{align}  
where $\textsf{P}_n$ and $\textsf{P}_{n\cdot m}$ are universal polynomials
specifying the lambda ring structure. From the series
$\lambda_t(1)^m=(1+t)^m=\sum {m\choose r} t^r$, we see that the binomial
coefficients are actually lambda operations on the ring of integers. The
binomial coefficients can be obtained as specializations of the elementary
symmetric functions of $m$ variables at $x_i=1$. More explicitly, this reads
for a formal alphabet $X=x_1+x_2+x_3+\ldots$
\begin{align}\label{lambdaE}
\lambda_t(X)&=\sum_{i\ge 0} \lambda_t(1+x_i t) =
1 +\lambda^1(x)t + \lambda^2(x)t^2+ \ldots \nn
&= 1 + (x_1+x_2+x_3+\ldots)t + 
(x_1x_2+x_1x_3+\ldots x_2x_3+x_2x_4+\ldots )t^3 \ldots \nn
&= \sum_{n=0}^\infty e_n(x) t^n
\end{align}
The polynomial $\textsf{P}_n$ is derived from the inverse Cauchy kernel
$\lambda_t(xy)=\prod_{i,j} (1-x_iy_jt)$ and the second polynomial 
$\textsf{P}_{n\cdot m}$ is obtained by expanding 
$\lambda_t(\lambda^q(x)) = \prod_i (1-(x_{i_1}\ldots x_{i_q})t)$
where the indices run in $1\le i_1<\ldots<i_q\le n$ and the inductive limit
$n\rightarrow \infty$ is understood.
\mybenv{Example}
Consider the following two binomial identities which emerge from the lambda
structure 
\begin{align}
\lambda^2(xy) 
&= {xy\choose 2} = x^2{y\choose 2} + y^2{x\choose 2}
   -2{x\choose 2}{y\choose 2} \nn
&= x^2 \lambda^2(y) + y^2\lambda^2(x) -2 \lambda(x)\lambda(y)\nn
\lambda^2(\lambda^2(x)) &= {{x\choose 2}\choose 2} 
 ={x\choose 3}{x\choose 1}-{x\choose 4} \nn
&=\lambda^3(x)\lambda^1(x)-\lambda^4(x)
\end{align}
\myeenv
While in the Hopf algebraic treatment of series one acts on coefficients 
(ring elements), the lambda structure lifts this to the ring $1+R\PS{t}^+$
of formal power series. This motivates our simultaneous treatment of series
and generating functions in the additive and multiplicative case. Lambda 
rings contain information about the representation theory of the group
structure at hand, in particular the ring of symmetric functions is a special
lambda ring and contains information about the representation theory of the
general linear group and the symmetric group. From our point of view, we
should keep in mind that primitive elements are related with the Adams
operations employed for example in $K$-theory. The Adams operations $\Psi$ 
actually also define a lambda ring structure, but in general not vice versa. 
The $\Psi$-ring is then isomorphic to a lambda-ring. Definition and
compatibility with the lambda operations is given for $x,y\in\bfR$ in the 
lambda ring $1+R\PS{t}^+$ as
\begin{align}
\Psi^1(x) &= x \nn
\Psi^n(1) &= 1 \nn
\Psi^n(x+y) &= \Psi^n(x) + \Psi^n(y) \nn
\Psi^n(xy) &= \Psi^n(x)\Psi^n(y) \nn
\Psi^n(\lambda^n(x)) &= \lambda^n(\Psi^n(x)) \nn
\Psi^n(\Psi^m(x)) &= \Psi^{n\cdot m}(x) = \Psi^m(\Psi^n x)) 
\end{align}
The last equation defines the operation of plethysm or composition of
representations. The relation between lambda and Adams operations is 
given by
\begin{align}
\dt\log(\lambda_t(x)) 
&= \sum_{n\ge 0}(-1)^n \Psi^{n+1}(x)t^n
\end{align}
A lambda ring element is called binomial if it fulfils
$\lambda_t(x)=(1+x)^a$. It turns out, that an element is binomial if
$\Psi^n(x)=x$ for all $n\in\openZ$, hence for our integers we find
$\Psi^n(m)=m$ and all integers are binomial. However, the Adams operations
will act nontrivially on the ring of formal power series over $\openZ$.

Recalling a standard construction, we define an $\bfR^\omega$ ring as the set
of countable sequences $[r_1,r_2,r_3,\ldots]$ with addition \textit{and} 
multiplication defined component wise. Define for every integer the Adams
operations on $\bfR^\omega$ as $\Psi^n([r_1,r_2,r_3,\ldots]) = 
[r_n,r_{2n},r_{3n},\ldots]$. It can be shown that $\bfR^\omega$ is a 
$\Psi$-ring. Now, one can define a morphism $L$ relating the lambda ring 
$1+\bfR\PS{t}^+$ with $\bfR^\omega$ such that 
\begin{align}
L\circ \lambda_t &= \Psi
\end{align}
$L$ can be defined as $\dt\log(1+e_1t+e_2t^2+\ldots) = \sum
(-1)^nr_{n+1}t^n$. In characteristic zero an inverse exists 
$L^{-1} : \bfR^\omega\rightarrow 1+\bfR\PS{t}^+$ and can be given 
by\footnote{%
We do just mention that this is the place to think of moment-cumulant
relations and Spitzer's identity \cite{rota:1969a,rota:1969b}.
A few more glimpses will come up in section \ref{sec:Occupation}.}
\begin{align}
L^{-1}([b_1,b_2,b_3,\ldots]) &= \exp(-g(t)) \nn
g(t) &= \sum_{n+1}(-1)^n b_n t^n,
\end{align} 
see \cite[page51]{knutson:1973a}. Besides the fact that the map $L$ plays an 
important role in the description of the $K$-theory of central functions on 
$G$ with values in $K$, we are interested in the case where $L$ is related
to universal Witt rings.

Let $W_\bfR$ be the set of all elements $[w_1,w_2,w_3,\ldots]$, $w_i\in \bfR$. 
Hence as sets $W_\bfR=\bfR^\omega$. We put a new ring structure on $W_\bfR$ 
defined by a map $M : W_\bfR \rightarrow \bfR^\omega$ which is given as
\begin{align}
M([w_1,w_2,w_3,\ldots]) &= [r_1,r_2,r_3,\ldots] \nn
r_n &= \sum_{d\mid n} d\, w_d^{\frac{n}{d}}
\end{align} 
\mybenv{Example}
\begin{align}\label{ex:rw}
r_1 &= w_1,\quad r_2= w_1^2+2w_2,\quad r_3=w_1^3+3w_3 \nn
r_4 &= w_1^4+2w_2^2+4w_1,\quad r_5=w_1^5+5w_5,\quad\ldots
\end{align}
\myeenv
The set $W_\bfR$ is closed under sum and product in $\bfR^\omega$ which 
establishes the ring structure. $W_\bfR$ is called Witt ring. In fact there
are universal polynomials $F_i,G_j$ with integer coefficients such that 
\begin{align}
M([w_1,w_2,w_3,\ldots]+[v_1,v_2,v_3,\ldots]) 
&= M([F_1(w_1,v_1),F_2(w_1,w_2,v_1,v_2),\ldots]) \nn
M([w_1,w_2,w_3,\ldots].[v_1,v_2,v_3,\ldots]) 
&= M([G_1(w_1,v_1),G_2(w_1,w_2,v_1,v_2),\ldots])
\end{align}
and the variables in the $F_i,G_i$ run in the set of variables $w_d,v_l$ 
where $d,l$ are divisors of $i$. The functor which relates these rings is
called the Witt functor.

We can model the basis change from the $w$ into the $r$ basis by using our
coproduct of multiplication. We define an action of $M$ on $w$ by means of 
multiples of `Adams operations' as
\begin{align}
M (w_n) &= \sum_{d\mid n}d\cdot [d\mid w]^{\frac{n}{d}} 
\end{align}
Note that after the universal polynomials have been computed they can be used
to define a ring structure on arbitrary rings, torsion free or not (hence in
any characteristic). The explicit construction is as follows. First construct
a lambda map $f : W_\bfR \rightarrow 1+\bfR\PS{t}^+$, $f([w])=\prod(1+w_d(-t)^d)$ 
and then apply the map $L$
\begin{align}
L\circ f([w]) 
&= \dt\log\prod_d (1-w_d(-t)^d) \nn
&= \sum_d \frac{dw_d(-t)^{d-1}}{1-w_d(-t)^d} \nn
&= \sum_d \frac{-dw_d(-t)^{d-1}}{t}(1+w_d(-t)^d+w_d(-t)^{2d}+\ldots) \nn
&= \sum_d (-1)^{n+1}\big(\sum_{d\mid n} dw_d^{\frac{n}{d}}\big)t^{n-1} \nn
&= \sum_d (-1)^{n+1} r_n t^{n-1}
\end{align} 
From the expansion of $f(w)$ one obtains an explicit relation between the
elementary symmetric functions $e_n$ and the $w_n$, compare with eqn. 
(\ref{ex:rw}), as well as with eqn. (\ref{lambdaE})
\begin{align}
\sum_n e_n t^n 
&= f(w) = \prod_d (1-w_d(-t)^d) \nn
&= (1+w_1t)(1-w_2t^2)(1+w_3t^3)(1-w_4t^4)\ldots \nn
&= 1+(w_1)t +(-w_2)t^2+(w_3-w_1w_2)t^3+(-w_4+w_1w_3)t^4+\ldots,
\end{align}
which can be solved recursively for the $w_n$. One obtains
\begin{align}
w_1 &= e_1,\quad w_2=-e_2,\quad w_3=e_3+e_1e_2,\quad
w_4  = -e_4+e_3e_1+e_2e_1^2 \nn
w_5 &= e_5-e_4e_1-e_3e_2-e_1e_2^2+e_1^2e_3+e_1^3e_2,\quad\ldots
\end{align}
These polynomials play a major role in combinatorics of necklaces, and in 
the study of the Burnside ring 
\cite{dress:siebeneicher:1988a,scharf:thibon:1996a}. Our interest is in 
employing these techniques in the theory of plethysms, as will be shown 
elsewhere.

\section{\label{sec:Applications}Applications:}
\setcounter{equation}{0}\setcounter{mycnt}{0}

Before we turn to the applications, we should remark that the presented
constructions are universal in a categorial sense. Hence one can have the
hope to employ these methods in seemingly remote mathematical fields as
long as the same functorial relations hold. It is hence more that a hope
that such elementary structures will arise again in operator algebras of
quantum fields, or in representation theoretic issues.

\subsection{\label{sec:SymFct}Symmetric function theory}

Consider the ring of symmetric functions $\Lambda$, as described for example
in \cite{macdonald:1979a}. This ring is graded by degree 
$\Lambda=\oplus_{n\ge 0}\Lambda^n$. The ring structure is given by addition of
polynomials and multiplication of polynomials in the usual fashion. A basis 
for this ring is indexed by partitions of integers. Let
$\lambda=(\lambda_1,\ldots,\lambda_k)$ with parts $\lambda_i$ such that
$\lambda_1\ge\ldots\ge\lambda_k$ and $\sum_{r=0}^k\lambda_r=n$. One can give a
multiset description by counting the multiplicity of the parts as
$\lambda=[1^{r_1}2^{r_2}3^{r_2}\ldots n^{r_n}\ldots]$. The $r_i$ give the
number of parts $\lambda_j=i$. Standard bases of the ring of symmetric
functions are the elementary symmetric functions $e_\lambda$, the complete
symmetric functions $h_\lambda$ the power sum basis $p_\lambda$, the Schur
function basis $s_\lambda$ and the monomial symmetric functions $m_\lambda$
among others. For our interest, we want to concentrate on the monomial
symmetric functions. These functions can be defined for any partition
$\lambda$ as (we use the variable or later alphabet $a$ from now on to comply
with \cite{rota:stein:1994a,rota:stein:1994b})

\begin{align}
m_\lambda &=
\sum_{i_1< \ldots <i_{r_1}}
\sum_{j_1< \ldots <j_{r_2}} \ldots
\sum_{l_1< \ldots <l_{r_k}}
a_{i_1} \ldots a_{i_{r_1}}
a^2_{j_1} \ldots a^2_{j_{r_2}} \ldots
a^k_{l_1} \ldots a^k_{l_{r_k}}
\end{align}
where all indices $i_s,j_t,\ldots$ are distinct. With Rota-Stein
\cite{rota:stein:1994a,rota:stein:1994b} we abbreviate this as
\begin{align}\label{monomialM}
m_\lambda &=(a^1)^{(r_1)}(a^2)^{(r_2)}\ldots (a^k)^{(r_k)} 
\end{align} 
Where we made the following identification 
\begin{align}
(a^i)^{(r_i)} 
&= \sum a^i_{j_1} \ldots a^i_{j_{r_i}},\quad j_{i_1}<\ldots< j_{r_i}
\end{align}
and the $a$ is now from a formal alphabet $A$, also called letter.
Such a monomial is an element of the space $\pleth[A]$ which is defined 
in the following way: Let $A$ be an alphabet (formal variable). Construct
the symmetric tensor algebra over $A$ as $\tens[A]$ spanned by monomials
$(a^n)$. Again $\tens[A]^+$ is the augmentation ideal with zero constant 
part. This is the module underlying the algebra of formal power series 
(in the commuting variables $a^1,a^2,a^3,\ldots$). Then form the divided
powers algebra over this \textit{module}. This module is generated by 
$(a^n)^{(k)}$ for all $n\ge 1$, $k\ge 0$. Hence the module underlying 
$\pleth[A]$ is $\divp\tens[A]^+$. The algebra structure is given by the 
divided power algebra rules for the indices in parentheses. We want to 
induce, following Rota and Stein \cite{rota:stein:1994a,rota:stein:1994b}, 
a new  multiplicative structure on this module in such a way that the
monomials $m_\lambda$ multiply in the same fashion as the monomial 
symmetric functions. This is done using a cliffordization as follows:
Recall that the $[r_1,r_2,r_3,\ldots]$ are the multiplicities of the
parts $\lambda_i$ in $\lambda$. The coproduct of an element $m_\lambda$
in $\pleth[A]$ is given as
\begin{align}
\DM(m_\lambda) 
&= m_{\lambda[1]}\otimes m_{\lambda[2]}\nn
&=\sum_{r_i^\prime+r_i^{\pprime}=r_i,~\forall i} 
(a^1)^{(r_1^\prime)}(a^2)^{(r_2^\prime)}\ldots(a^k)^{(r_k^\prime)}
\otimes
(a^1)^{(r_1^\prime)}(a^2)^{(r_2^{\pprime})}\ldots(a^k)^{(r_k^{\pprime})}\nn
&= \delta^\DP(
(a^1)^{(r_1)}(a^2)^{(r_2)}\ldots(a^k)^{(r_k)})
\end{align}
as in the case of the unrenormalized coproduct. The product is defined
for the divided powers component wise on the monomials as
\begin{align}
(a^i)^{(r_i)} \cdot (a^j)^{(r_j)}
&= (a^i)^{(r_i)} (a^j)^{(r_j)}\quad i\not=j\nn 
(a^i)^{(r_i)} \cdot (a^i)^{(s_i)} &= {r_i+s_i\choose r_i} (a^i)^{(r_i+s_i)}
\end{align}
The next step is to define a pairing, sic a bilinear form. This pairing
is assumed to be a Laplace pairing, which literally provides an expansion
formula in terms of coproducts, see
\cite{rota:stein:1994a,fauser:2002c,brouder:fauser:frabetti:oeckl:2002a}.
The particular pairing will be a 2-cocycle in terms of the cohomology used
above eqn. (\ref{coboundary}), which guarantees that the deformed product
is associative. One sets:
\begin{align}
\langle (a^i)^{(r)}\mid (a^j)^{(s)}\rangle
&= \delta_{r,s}(a^{i+j})^{(s)}
\end{align}
extended by bilinearity. Actually this pairing is used to reintroduce the 
algebra structure of formal power series in the `inner space' $\tens[A]^+$.
Using this pairing we introduce the Drinfeld twisted or star product which
we call with Rota-Stein a \textbf{Clifford product} or circle product in the 
usual way
\begin{align}
\label{circleprod}
m_\lambda \circ m_\mu 
&= \langle m_{\lambda(1)}\mid m_{\mu(1)} \rangle m_{\lambda(2)}m_{\mu(2)}\nn
&= \textsf{R}^\pi_{\lambda,\mu} m_\pi
\end{align}
where the $\textsf{R}^\pi_{\lambda,\mu}$ are the structure constants of
this algebra in the monomial symmetric function basis. In fact Rota and
Stein showed that this process comes up with the same algebraic structure
as the monomial symmetric functions have under the point wise product.
It is noteworthy to state, that the Clifford product gives a direct route
to the \textbf{Littlewood-Richardson coefficients} 
$\textsf{C}^\pi_{\lambda,\mu}$ of Schur function multiplication. A basis
change from monomial symmetric functions to Schur functions is given by
the Kostka matrix $K_{\lambda,\mu}$, which can be combinatorially obtained,
see \cite{macdonald:1979a}. Hence one finds
\begin{align}
s_\lambda &= K_\lambda^\mu m_\mu \nn
s_\lambda\circ s_\mu
&= K_\lambda^{\lambda^\prime}K_\mu^{\mu^\prime}
  \textsf{R}^{\pi^\prime}_{\lambda^\prime,\mu^\prime}
  (K^{-1})^{\pi}_{\pi^\prime} s_\pi
 = \textsf{C}^{\pi}_{\lambda,\mu} \, s_\pi
\end{align}
This exhibits that the circle product is the ordinary product of polynomials
and symmetric functions as claimed. We can summarize this as
\mybenv{Rota-Stein Theorem} 
The module $\divp\tens[A]^+$ underlying the plethystic Hopf algebra
together with the circle product (\ref{circleprod}) forms the symmetric 
function algebra (in monomial basis).
\myeenv
This result demonstrates in a rather intriguing fashion, via Hopf algebraic
mechanisms, the combinatorial origins of symmetric function theory.
\mybenv{Example}
Let $m_{1}=(a^1)^{(1)}$ and compute
\begin{align}
m_{1}\circ m_{1}
&= (a^1)^{(1)}\circ (a^1)^{(1)} \nn
&= \sum_{r=0}^{1} \sum_{s=0}^{1} 
\langle (a^1)^{(r)}\mid (a^1)^{(s)} \rangle (a^1)^{(1-r)}(a^1)^{(1-s)} \nn
&= \sum_{r=0}^{1} \sum_{s=0}^{1} 
\delta_{r,s} (a^2)^{(s)} (a^1)^{(1-r)}(a^1)^{(1-s)} \nn
&=\sum_{r=0}^{1}
(a^2)^{(r)} (a^1)^{(1-r)}(a^1)^{(1-r)} \nn
&= (a^2)^{(1)}(a^1)^{(0)}(a^1)^{(0)}
 + (a^2)^{(0)}(a^1)^{(1)}(a^1)^{(1)} \nn
&= 2 (a^1)^{(2)}+ (a^2)^{(1)} \nn
&=  2 m_{11}+m_{2}
\end{align}
A more complicated case is
\begin{align}
m_{5}\circ m_{22} &= (a^5)^{(1)}\circ (a^2)^{(2)} \nn
&= 
 \langle (a^5)^{(0)}\mid (a^2)^{(0)}\rangle (a^5)^{(1)}(a^2)^{(2)}
+\langle (a^5)^{(1)}\mid (a^2)^{(1)}\rangle (a^5)^{(0)}(a^2)^{(1)} \nn
&= (a^7)^{(0)}(a^5)^{(1)}(a^2)^{(2)} 
 + (a^7)^{(1)}(a^5)^{(0)}(a^2)^{(1)} \nn
&= m_{522} + m_{72}
\end{align}
And checking a result of Rota and Stein we get
\begin{align}
m_{111}\circ m_{11} &= (a^1)^{(3)}\circ (a^1)^{(2)} \nn
&=
 \langle (a^1)^{(2)}\mid (a^1)^{(2)}\rangle (a^1)^{(1)}(a^1)^{(0)}\nn
&+\langle (a^1)^{(1)}\mid (a^1)^{(1)}\rangle (a^1)^{(2)}(a^1)^{(1)}\nn
&+\langle (a^1)^{(0)}\mid (a^1)^{(0)}\rangle (a^1)^{(3)}(a^1)^{(2)}\nn
&= (a^2)^{(2)}(a^1)^{(1)}(a^1)^{(0)} 
 + (a^2)^{(1)}(a^1)^{(1)}(a^1)^{(1)} 
 + (a^2)^{(0)}(a^1)^{(3)}(a^1)^{(2)}  \nn
&= (a^2)^{(2)}(a^1)^{(1)} 
 + 2 (a^2)^{(1)}(a^1)^{(2)} 
 + {3+2\choose 2} (a^1)^{(5)}  \nn
&= m_{221}+2m_{211}+10m_{11111}
\end{align}
\myeenv
Since we know that the deformation yields an associative product it 
follows that the pairing introduced on $\pleth[A]=\divp\tens^+$ is a
2-cocycle. There is hence a chance that this 2-cocycle 
$\langle .\mid .\rangle$ is actually derived from a 1-cochain $\eta$, 
giving $\langle .\mid .\rangle=(\partial_2\eta^{-1})(.,.)$, since we know 
that the actual product of monomial symmetric functions is isomorphic to the 
original products of polynomials in the variable(s) $A$. Indeed, Rota and 
Stein loc. cit. provide the map $\eta : \tens[A]^+\rightarrow \pleth[A]\cong 
\divp\tens[A]^+$ such that
\begin{align}
\eta((a)^{(n)}) &= \sum_{\lambda\vdash n} 
(a^1)^{(r_1)}(a^2)^{(r_2)}(a^3)^{(r_3)}\ldots (a^k)^{(r_k)}
\end{align}
where the sum is over all partitions $\lambda$ of $n$. This resembles a
classical identity of symmetric functions
\begin{align}
h_n &= \sum_{\lambda\vdash n} m_\lambda
\end{align}
The complete symmetric functions and monomial symmetric functions form
mutually dual bases $\langle h_\lambda\mid m_\mu\rangle = 
\delta_{\lambda,\mu}$ The Clifford product and the original product on
$\pleth[A]$ are hence related by
\begin{align}
\eta( (a^i)^{(r_i)} ) \circ \eta( (a^j)^{(r_j)} )
&= 
\eta( (a^i)^{(r_i)}(a^j)^{(r_j)} )
\end{align}
compare with eqn. (\ref{circleprod}). Inspection shows that the scalarproduct
introduced on $\pleth[A]$ is related to the convolutive inverse 
$\eta^{-1}$ of $\eta$ as
\begin{align}
\langle (a^i)^{(r_i)} \mid (a^j)^{(r_j)} \rangle 
&= (\partial \eta^{-1}) ((a^i)^{(r_i)},(a^j)^{(r_j)}) 
\end{align}
We close this discussion by saying that the $(a)^{(n)}$ have been refereed to
by Rota and Stein as complete symmetric functions in the inverse alphabet
$A^{\#}$ under the  map $\eta$, explicitly $h_n = \eta((a^{\#})^{(n)})$. 
A detailed exposition of these facts is beyond the scope of the present
paper, and is postponed to another paper. Also we will not go into the details
of plethysms, which is a major point of interest in the present development,
but see \cite{fauser:jarvis:king:wybourne:2004a,fauser:jarvis:2005a}.

\subsection{\label{sec:Occupation}Occupation number representations}

Before we try to match our results to quantum field theory, we give a 
short exposition of occupation number states in quantum mechanics. For
the sake of simplicity we will use one label type, quanta of type $a$,
and a bosonic scenario. The discussion will also be insensitive to the basis
change between $p,q$ and $a,a^\dagger$. Physically some of the algebras need
to be interpreted in the $p,q$-basis however.\footnote{%
One has to study the algebras $\openk\PS{q}\otimes \divp[p] \cong
\openk\PS{a}\otimes \divp[a^\dagger] \cong \openk\PS{t}[t^{-1}]$, where the 
last has to be interpreted as the localization of $\openk\PS{t}$ at 
$t=0$, and constitutes the quotient field of the power series ring.}
The precise setting will be outlined elsewhere.

\paragraph{Algebraic setup:}

Let ${\cal H}$ be a $L^2$ Hilbert space. We construct a countable
basis, the occupation number basis, for this space out of a cyclic vector
$\vert 0\rangle$ called the `vacuum'. Hence we demand that there is a non
nilpotent map $a^\dagger : {\cal H} \rightarrow {\cal H}$ such that 
\begin{align}
{\cal H} &= \oplus_n (a^\dagger)^n \vert 0\rangle = \oplus_n \vert n\rangle 
\end{align}  
Each state $\vert n \rangle$ is interpreted as having $n$ quanta of type $a$, 
while $\vert 0\rangle$ is the state having no such quanta. A general
state $\vert\psi\rangle$ of the Hilbert space ${\cal H}$ can be given as a
linear combination of the occupation number basis
\begin{align}
\vert\psi\rangle 
&= \sum_{n\ge0} \psi_n \vert n\rangle 
 = \sum_{n\ge0} \psi_n (a^\dagger)^n \vert 0\rangle  
\end{align}
Forgetting about convergence issues, one might reinterpret this as an 
element in the formal power series algebra generated by $a^\dagger$ with
coefficients in $\openC$, equivalently $\openC\PS{a^\dagger}$. The action of
$a^\dagger$ on an element $\vert n\rangle$ is given by
\begin{align}\label{occStates}
a^\dagger \vert n \rangle 
&= a^\dagger\,(a^\dagger)^n \vert 0\rangle 
 = (a^\dagger)^{n+1} \vert 0\rangle = \vert n+1\rangle, \nn
(a^\dagger)^n\,(a^\dagger)^m 
&= (a^\dagger)^{n+m}
\end{align}
where the last line makes the algebra product explicit. Of course, the
product of operators $a^\dagger$ amounts to an addition in the exponent.
Finally it should be noted that in eqn. (\ref{occStates}) no explicit
normalization factors are included. 

Quantum physics needs not only the creation of modes but also annihilation
of modes. This is done by the utilization of annihilation operators $a$.
It is commonly accepted that these operators fulfil the same type of algebra.
However we know from our discussion above (eqn. (\ref{SymDivp})), and as is
well known in mathematical literature, see e.g. \cite{weyman:2003a}, that the
dual of a formal power series algebra (polynomial algebra) is a divided powers
algebra. Hence we deal with two types of structures here which are related
as in the following diagram\footnote{%
In fact we are dealing here with a pair of dual Hopf algebras, hence with a
Drinfeld quantum double.}

\begin{align}
\label{diag1}
\begin{array}{c@{\hskip 3truecm}c}
\Rnode{A}{\openk\PS{a}} & \Rnode{B}{\openk{\PS{a^\dagger}}}\\[2truecm]
\Rnode{C}{\divp[a]} & \Rnode{D}{\divp[a^\dagger]}
\end{array}
\ncline{<->}{A}{B}\Aput{\text{herm. conj.}}
\ncline{<->}{C}{D}\Bput{\text{herm. conj.}}
\ncline{<->}{A}{C}\Aput{\text{duality}}
\ncline{<->}{B}{D}\Aput{\text{duality}}
\ncline[linestyle=dotted]{<->}{A}{D}\Aput{\text{QM}}
\ncline[linestyle=dotted]{<->}{B}{C}\Aput{\text{QM$^*$}}
\end{align}

That quantum mechanics (QM) resides on the diagonals will become clearer 
below. QM$^*$ is the version of QM where $p$ and $q$ or $a$ and $a^\dagger$
are interchanged in their meaning. In characteristic zero, the algebras 
$\openk\PS{x}$ and $\divp[x]$ are related by an isomorphism. Of course, the
complex number ring $\openC$ is of characteristic zero and such an isomorphism
is at our disposal, but, our point is that one looses insight into the
combinatorial setting of QM by adopting it implicitly, as is usually done. 
Moreover, a possible generalization to finite characteristic is prevented.

Dropping habitual reflexes, it is at a first glance not totally obvious how to
relate annihilation and creation operators. To implement this relation we
introduce a duality. Hence we define linear forms as
\begin{align}
\eval(f_{a^{\dagger n}} \otimes a^{\dagger m}) &= \delta_{n,m}
\end{align}
We know by now that the algebra of the duals is a divided powers algebra.
For notational convenience we adopt $f=f_1=f_{a^\dagger}$, 
and $f_n=f_{a^{\dagger n}}$, and demand that $f_1$ 
generates\footnote{
Strictly speaking, the $f_i$ cannot be multiplicatively generated in finite
characteristic due to the numerical factors appearing. This can be overcome by
the usage of a Baxter operator $R(f_i)=f_{i+1}$. A divided powers algebra 
together with this operator form a Rota-Baxter algebra of weight $0$, see 
also page \pageref{baxter}.}
the divided powers algebra $\divp[f]$
\begin{align}
n! f_n &= (n-1)! f_1 f_{n-1} = f_1^n\nn
f_n f_m &= {n+m\choose n} f_{n+m}
\end{align}
We can now introduce dual states ${\cal H}^* = \hom({\cal H},\openk)$ as
\begin{align}
\langle 0 \vert f_n 
&= \langle n\vert \nn
\langle\psi\vert 
&= \sum_n \psi_n \langle n\vert = \sum_n \psi_n \langle 0\vert f_n
\end{align}
Usually operators $a$ are introduced which fulfil the \textbf{same} type of
power series algebra $\openk\PS{a}$ as the creation operators assuming the
isomorphism $f_i \cong a^i/i!$. The multiplication law of the $a^i$ becomes
then $a^i\, a^j = a^{i+j}$. The evaluation map $\eval :
\openk[a]\otimes \openk\PS{a^\dagger} \rightarrow \openk$ allows then to 
introduce a coalgebra structure on the ordinary power series algebra 
$\openk\PS{a^\dagger}$ relating to standard notation of QM
\begin{align}
\DP((a^\dagger)^0) 
&=\DP(1) = 1 \otimes 1 \nn
\DP(a^\dagger) 
&=a^\dagger\otimes 1 + 1 \otimes a^\dagger \nn
\DP((a^\dagger)^n) 
&= \sum_{r=0}^n (a^\dagger)^r \otimes (a^\dagger)^{n-r} 
\end{align}
This is the \textbf{unrenormalized coproduct of addition}, indexing the 
$a$-mode creation operator $a^\dagger$. However, doing so destroys the duality 
between these two algebras, as we have seen studying the additive Hopf
convolution. This forces one to introduce \textbf{normalization factors}
which are artificial. The linear forms $f_n$ can be described by 1-cochains
on this algebra. Especially we can assign (the action of) $f_1$ to the 
annihilation operator $a$ as the 'name'\footnote{
The concept of a 'name' of a map stems from category theory
\cite{lawvere:rosebrugh:2003a}. One might however stick to a more mechanical picture. 
Let $f_1,f_2,\ldots$ be the buttons on a calculator, the enter key the 
evaluation map, than strictly the button is (tagged by) the name of the 
function you want to use. Hence $a$ is the name of the function $f_1$ acting
by evaluation (Kronecker pairing).}
of the linear form with action given by evaluation. For an alternative development of
generalized (boson) algebras viewed as Hopf algebras, see
\cite{tsohantis:paolucci:jarvis:1997a}.

\paragraph{Branchings with respect to polynomial algebra and divided powers:}

Let us study with more care contraction maps, as developed is section 
\ref{sec:2-3}, \ref{sec:2-4}, \ref{sec:3-4} and, \ref{sec:3-5},
related to branching operators, starting with division $\JJ{/\div}$
\begin{align}
\JJ{/\div}(a \otimes (a^\dagger)^n) 
&= (\eval\otimes \Id)(\Id\otimes \DP)(K \otimes \Id)
    (a \otimes (a^\dagger)^n)\nn
&= (\eval \otimes \Id)(\Id\otimes \DP)
    (f_1 \otimes (a^\dagger)^n)\nn
&= \sum_{r=0}^n f_1((a^\dagger)^r) \otimes (a^\dagger)^{n-r} \nn
&= (a^\dagger)^{n-1} 
\end{align}
Using the unrenormalized coproduct we note that $a\JJ{/\div}$ \textbf{fails}
to be a derivation, and as an operator would therefore \textbf{commute} with 
$a^\dagger$. Dualizing now correctly the divided powers algebra, we have to
use the renormalized coproduct of addition here.
\begin{align}\label{eqn:5-25}
a( (a^\dagger)^n ) 
&=\underline{\JJ{\relax}}_{/\div}(a \otimes (a^\dagger)^n) \nn
&= (\eval\otimes \Id)(\Id\otimes \DPR)(K \otimes \Id)
    (a \otimes (a^\dagger)^n)\nn
&= (\eval\otimes \Id)(\Id\otimes \DPR)
    (f_1 \otimes (a^\dagger)^n)\nn
&= \sum_{r=0}^n {n\choose r} f_1((a^\dagger)^r) \otimes (a^\dagger)^{n-r} \nn
&= n (a^\dagger)^{n-1} 
\end{align}
Hence the dualized elements give rise to `names' $a$ which \textbf{act
as derivations}. In fact we proved already earlier that these branchings 
fulfil the Leibniz rule. 

\paragraph{Canonical commutation relations:}

Using the branching operators more systematically we will now show how the
two coproducts, renormalized and unrenormalized intertwine to build the
core features of QM. We employ Kronecker duality, a power series algebra 
structure for the $a^{\dagger n}$ and a divided powers algebra for the
duals $f_n$. This allows us to compute immediately the 
\textbf{canonical commutation relations} 
\begin{align}
a\, a^\dagger\,(a^\dagger)^n  -a^\dagger\, a\,(a^\dagger)^n 
& = \Id (a^\dagger)^n \nn
a (a^\dagger)^{n+1} - a^\dagger a (a^\dagger)^n &=
(n+1)(a^\dagger)^n -n (a^\dagger)^n = 1\,(a^\dagger)^n \nn
{\text{so that}\quad}[a,a^\dagger]_{}&= \Id 
\end{align}
Since this is a major point of our development, we will give a Hopf 
version of this calculation too. We use the following Sweedler index 
notation $\DP(n)=n_{(1)}\otimes n_{(2)}$ and 
$\DPR(n)=n_{[1]}\otimes n_{[2]}$ and compute the action of $a^n$ on the 
product of $a^{\dagger p} a^{\dagger q}$ for arbitrary $m=p+q$. The general 
formula, see \cite{fauser:2002c} for a graded version, is given as
\begin{align}
\underline{\JJ{\relax}}_{/\div}(a^n\otimes (a^{\dagger p} a^{\dagger q}))
&=f_n( a^{\dagger p[1]}a^{\dagger q[1]})
  a^{\dagger p[2]} a^{\dagger q[2]} \nn
&=f_{n(1)}(a^{\dagger p[1]})f_{n(2)}(a^{\dagger q[1]})
  a^{\dagger p[2]} a^{\dagger q[2]} 
\end{align}
which specializes for $m=n+1$, $p=1$, $q=n$ to
\begin{align}
\underline{\JJ{\relax}}_{/\div}(a\otimes(a^{\dagger 1} a^{\dagger n}))
&= \sum_{{p^\prime+p^\pprime=1}\atop{n^\prime+n^\pprime=n}}
  {p^\prime+p^\pprime\choose 1}{n^\prime+n^\pprime\choose n}
  f_1(a^{\dagger p^\prime})f_0(a^{\dagger n^\prime})
  a^{\dagger p^\pprime}a^{\dagger n^\pprime} \nn
&~~~~~ + {p^\prime+p^\pprime\choose 1}{n^\prime+n^\pprime\choose n}
  f_0(a^{\dagger p^\prime})f_1(a^{\dagger n^\prime})
  a^{\dagger p^\pprime}a^{\dagger n^\pprime} \nn
&= 1\cdot 1\cdot a^{\dagger n} + 1\cdot (n-1) \cdot a^{\dagger} a^{\dagger (n-1)} 
\end{align}
as in eqn.(\ref{eqn:5-25}). This process does \textbf{only} work due to the
usage of the \textbf{unrenormalized coproduct of addition} dualized from the
divided powers product of the duals, since only this pair of algebraic
structures fulfil the homomorphism axiom which allows Laplace expansions.

We are now able to compute the scalar product between two states.
Note the usage of \textbf{two different coproducts}:
\begin{align}\label{eqn:5-29}
\langle n &\mid m\rangle
 = f_n (a^\dagger)^m  
 = \langle 0 \vert \frac{1}{n!} a_n (a^\dagger)^m \vert 0 \rangle \nn
&= \frac{1}{n!} \sum\sum a_{n(1)}(a^{\dagger m[1]})a_{n(2)}(a^{\dagger m[2]}) 
 = (\DP^{(n-1)}(f_n)(\DPR^{(m-1)}(a^{\dagger m}))  \nn
&= \frac{1}{n!} 
\sum_{s_1=0}^m \sum_{s_2=0}^{s_1} \ldots \sum_{s_{n-1}=0}^{s_{n-2}}
{m\choose s_1}{s_1\choose s_2} \ldots {s_{n-2}\choose s_{n-1}} \nn
&~~~~~~~~~~~~~~~~~~~~~~
\langle 0 \vert a((a^\dagger)^{m-s_1}) a((a^\dagger)^{s_1-s_2}) \ldots 
a((a^\dagger)^{s_{n-2}-s_{n-1}}) \nn
&=\delta_{n,m} \frac{n(n-1)(n-2)\ldots 1}{n!} \langle 0 \vert 0 \rangle   \nn
&=\delta_{n,m}
\end{align}
We thus come up with properly normalized states, even without using
the common normalization using square roots of $n$. Evaluating the
coproducts directly in $\divp[f]\otimes \openk\PS{a^\dagger}$ does not 
need the fractions at all, and could be generalized to finite characteristic. 
The computation also sheds some light on the diagram (\ref{diag1}), explaining
why QM is sitting on the diagonal line. We want to emphasize the similarity
between this calculation and $S$-matrix calculations in quantum field 
theory (of course here $S=\Id$, look at line 3 of eqn. (\ref{eqn:5-29})). 
From the above calculation it is clear that we can deduce expectation values
of all operators which are described by creation and annihilation operators.

\paragraph{Normal ordering and Rota-Baxter operators:}

The divided powers algebra is a Baxter algebra of weight $\lambda=0$, that 
is a commutative algebra $\divp$ together with an Rota-Baxter operator 
$R : \divp \rightarrow \divp$ such that the following identity holds 
with $\lambda=0$
\begin{align}\label{baxter}
R(R(x)y +xR(y) +\lambda R(xy)) &= R(x)R(y)
\end{align}
This identity plays a prominent role in many areas of mathematics,\footnote{%
It is instructive to read the original papers by Baxter
\cite{baxter:1960a} Cartier \cite{cartier:1972a} and Rota with collaborators
\cite{rota:1969a,rota:1969b,rota:smith:1972a,rota:1995a}. Especially in the
last two references, one will find useful remarks about Spitzer's identity
and Warring's formula and their derivation from the `main theorem' of a free
Rota-Baxter algebra.}
notably in statistics and the path integral formalism of quantum field
theory \cite{thibon:luque:2002a}. For divided powers one finds that
\begin{align}
R(f_i) &= f_{i+1}
\end{align}
The Rota-Baxter operator is required to generate the $f$-basis of the dual
states, which cannot be multiplicatively generated in characteristic zero. 
The Baxter relation for $\lambda=0$ is the functional equation for integration 
and embodies the integration by parts formula. The Rota-Baxter operator can 
be used to introduce a deformed convolution and is deeply linked to Lie theory
and logarithmic functions \cite{bryant:2002a}. 

We will point out here only the relevance of Rota-Baxter operators to the 
problem of normal ordering and Stirling numbers of the second kind. Let 
$t$ be a formal variable of a polynomial series generating function, i.e. 
$t^n t^m = t^{n+m}$. Furthermore, let $t_{(n)}$ be the formal variables 
related to $t$ by $t_{(n)}= t(t-1)\ldots (t-n+1)$, the falling factorial. 
It seems to be known since the seventies that the normal ordering process 
of creation and annihilation operators provides another way to obtain
these numbers, for example see \cite{solomon:etal:2004a}, the results of 
Bender et al \cite{bender:brody:meister:1999a}, and also \cite{guo:2004a}.
Identifying $t=:a^\dagger a:$ one has
\begin{align}\label{Stirling2}
(:a^\dagger a: )^n 
&= t^n = \sum_{k=0}^n S(n,k) t_{(n)}
 = \sum_{k=0}^n S(n,k) :a^{\dagger k} a^k:\nn
S(n,k) &= \sum_{j=1}^k \frac{(-)^{k-j}j^n}{j!(k-j)!}
\end{align}
where the $S(n,k)$ are the Stirling number of the second kind. Let
${\cal D}$ be the forward difference operator ${\cal D}(t^n)=t^{n+1}-t^n$. 
Then one finds $k!S(n,k)= {\cal D}^k (t^n)\vert_{t=0}$. One can identify
$R(1)^n = t^n = (:a^{\dagger} a:)^n$ and $n!R^n(1)=n!R(\ldots R(1)\ldots)
= t_{(n)} = :a^{\dagger n} a^n:$. Iterating the Rota-Baxter identity, eqn. 
(\ref{baxter}), one can prove the `main theorem':
\mybenv{Theorem}\label{mainTheorem}
Let ($A,R$) be a commutative algebra and $R$ be a Rota-Baxter operator of 
weight 1. Let $R^k(1) = R(\ldots R(1)\ldots )$, $k$-iterations and 
$R(1)^k= R(1)\ldots R(1)$ $k$-factors. For $n\ge 1$ we have
\begin{align}
R(1)^n &= \sum_{k=1}^n k! S(n,k) R^k(1)
\end{align}
and 
\begin{align}
e^{R(1)u} 
&= \sum_{k=0}^n R(1)^n\frac{u^n}{n!} 
 = \sum_{n\ge 0}\sum_{k=0}^n 
   k!S(n,k)R^k(1) \frac{u^n}{n!}
\end{align}
is the exponential (divided powers) generating function for the Stirling
number of the second kind.
\myeenv

We will finish this subsection by rederiving such results showing that 
the mechanism developed for quantum field theory in 
\cite{fauser:2001b,brouder:fauser:frabetti:oeckl:2002a}
to describe the transition from time ordering to normal ordering and
vice versa, can be used in QM too if the combinatorially correct coproducts
are employed. Especially lookup the formulae (12) and (13) loc. cit. where 
all signs are plus signs since we are dealing with bosons. Let 
$W=\langle a^\dagger,a \rangle$ and consider the symmetric algebra
$\sym(W)=\oplus_n W^n =\sym(a,a^\dagger)$ spanned by normal ordered
monomials $:a^{\dagger n}a^m:$. $\sym(W)$ is graded. We combine the two
coproducts in a single notation
\begin{align}
\DH &=\left\{\begin{array}{cccc}
\delta^{\DP} & \text{if acting on~} a^{\dagger n} & \Rightarrow &
	\DH(a^{\dagger n})=\sum a^{\dagger r} \otimes a^{\dagger n-r} \\
\delta^{\DP} & \text{if acting on~} a^{n} & \Rightarrow &
	\DH(a^{n})=\sum a^{r} \otimes a^{n-r} \\
	\end{array}\right.
\end{align}
This is to say, that the Laplace pairing defined below is expanded with respect to the
renormalized coproducts. We define the Laplace pairing ${\cal F}$ as follows:
\begin{align}\label{LaplaceF}
{\cal F}(W^n,W^m) &=0\quad\text{if~} n\not=m, \nn
{\cal F}(1,1) &=1,\nn
{\cal F}(a,a) &=0,\quad {\cal F}(a,a^\dagger) =1,\quad
{\cal F}(a^\dagger,a) =0,\quad{\cal F}(a^\dagger,a^\dagger) =0,\nn
{\cal F}(u^\prime\cdot u^\pprime,v) 
&={\cal F}(u^\prime\otimes u^\pprime, \DH(v)), \nn
{\cal F}(u,v^\prime\cdot v^\pprime) 
&={\cal F}(\DH(u),v^\prime\otimes v^\pprime)
\end{align}
We can now expand the lhs. of the preceding theorem \ref{mainTheorem}
using a cliffordization with respect to the Laplace pairing eqn. 
(\ref{LaplaceF}), where in the cliffordization we use the unrenormalized
coproduct\footnote{To achieve normal ordering, it would be sufficient to use the
unrenormalized coproduct on the $a^\dagger$ algebra only.}. 
\begin{align}
(: a^{\dagger r} a^s:) \circ (: a^{\dagger m} a^n:)
&=
{\cal F}\Big(: a^{\dagger r[1]}a^{s[1]}:,
             : a^{\dagger m[1]}a^{n[1]}:
        \Big)
: a^{\dagger r[2]+m[2]} a^{s[2]+n[2]}:  
\end{align} 
It should be noted, that this process is not only applicable to
balanced terms having an equal number of creation and annihilation
operators, but applies generally. It can be shown, that the bilinear form
is zero unless one considers terms ${\cal F}(a^n,a^{\dagger m}) = 
\delta_{n,m}$, so two coproducts readily drop out.
\begin{align}
(: a^{\dagger r} a^s:) \circ (: a^{\dagger m} a^n:)
&=
{\cal F}\Big(: a^{s[1]}:,
             : a^{\dagger m[1]}:
        \Big)
: a^{\dagger r+m[2]} a^{s[2]+n}: \nn
&=: a^{\dagger r+m[2]} a^{s-m[1]+n}: 
\end{align} 

\mybenv{Example} We compute in detail a couple of products to show how
the normal ordering appears automatically as in \cite{fauser:2001b},
but this time with two different coproducts employed!
\begin{align}
a\circ a 
&=
{\cal F}(a,a)\, 1 + {\cal F}(1,1)\, :a^2: = :a^2:\nn
a^\dagger \circ a^\dagger 
&=
{\cal F}(a^\dagger,a^\dagger)\, 1 + {\cal F}(1,1)\, :a^{\dagger 2}:
 =\, :a^{\dagger 2}: \nn
a^\dagger \circ a 
&=
{\cal F}(a^\dagger ,a)\, 1 + {\cal F}(1,1)\, :a^{\dagger}a: 
 =\, :a^\dagger a: \nn
a\circ a^\dagger 
&=
{\cal F}(a,a^\dagger)\, 1 + {\cal F}(1,1)\, :a^\dagger a: 
 =\, :a^\dagger a: + 1 \nn
\end{align}
We obtain for $R(1)R(1)= :a^{\dagger}a:\circ :a^{\dagger}a:$
\begin{align}
:a^\dagger a: \circ :a^{\dagger} a: 
&=
({\cal F}\otimes \cdot)(\Id\otimes\textsf{sw}\otimes \Id)\nn
&~~~~~~
(a^\dagger a\otimes 1 + a^\dagger \otimes a + a\otimes a^\dagger + 1\otimes
 a^\dagger a) \nn
&~~~~
\otimes(a^\dagger a\otimes 1 + a^\dagger \otimes a + a\otimes a^\dagger + 1\otimes
 a^\dagger a) \nn
&~~~~~~
 {\cal F}(a,a^\dagger) :a^{\dagger} a: 
+{\cal F}(1,1) :a^{\dagger 2} a^2: \nn
&~~~~~~
 :a^{\dagger 2} a^2: + :a^{\dagger} a: 
\end{align} 
A more delicate computation is the following
\begin{align}
:a^\dagger a: \circ :a^{\dagger 2} a^2: 
&=
({\cal F}\otimes \cdot)(\Id\otimes\textsf{sw}\otimes \Id)\nn
&~~~~~~
(a^\dagger a\otimes 1 + a^\dagger \otimes a + a\otimes a^\dagger + 1\otimes
 a^\dagger a) \nn
&~~~~~~ \otimes(
 a^{\dagger 2}a^2\otimes 1 + 2a^\dagger a^2\otimes a^\dagger
 +a^2\otimes a^{\dagger 2} \nn
&~~~~~~~~~~~
 2a^{\dagger 2}a\otimes a + 4a^\dagger a\otimes a^\dagger a
 +2a\otimes a^{\dagger 2} a \nn
&~~~~~~~~~~~
 a^{\dagger 2}\otimes a^2 + 2a^\dagger \otimes a^\dagger a^2
 +1\otimes a^{\dagger 2} a^2) \nn
&= {\cal F}(1,1) :a^{\dagger 3} a^3: 
+ 2{\cal F}(a,a^\dagger) :a^{\dagger 2} a^2: \nn
&= :a^{\dagger 3} a^3: + 2\,:a^{\dagger 2} a^2:  
\end{align}
and similarly combining the above results one obtains
\begin{align}
:a^\dagger a:\circ :a^\dagger a:\circ :a^\dagger a:
&= :a^{\dagger 3} a^3: + 3\, :a^{\dagger 2} a^2: + :a^\dagger a:
\end{align}
In terms of Rota-Baxter operators using egn. (\ref{baxter}) this reads
as  follows
\begin{align}
R(1)^3 
&= 6R^3(1)+ 6R^2(1) + R(1)
 = [3!R^3(1)] + 2\cdot [2!R^2(1)] + [1!R^1(1)]
\end{align}
showing clearly that we can model the Rota-Baxter action using twisted
products induced by a Laplace pairing. Along similar lines we obtain
\begin{align}
:a:\circ :a^{\dagger n}: 
&=
\sum_{r=0}^n {\cal F}(a,a^{\dagger r})
\left(n\atop r\right):a^{\dagger n-r}:+ {\cal F}(1,1):a^{\dagger n} a: \nn
&= n :a^{\dagger n-1}: + :a^{\dagger n} a: \nn
:a^2:\circ :a^{\dagger n}: 
&=
\sum_{r=0}^n 
  {\cal F}(a^2,a^{\dagger r})\left(n\atop r\right):a^{\dagger n-r-2}:
 +{\cal F}(a,a^{\dagger r})\left(n\atop r\right)\left(2\atop 1\right):a^{\dagger n-r-1} a:
 +{\cal F}(1,1):a^{\dagger n-r} a^2: \nn
&=
  2\left(n\atop 2\right):a^{\dagger n-2}: 
 +2n:a^{\dagger n-1}a: + :a^{\dagger n}a^2: 
\end{align}
Commutation methods may seem to be computationally more efficient, however, we know
that software implementations  
\cite{ablamowicz:fauser:2002e,ablamowicz:fauser:2002d,ablamowicz:fauser:2002c}
of the Hopf algebraic procedure can be faster in many cases.
\myeenv

As a disclaimer, the reader should remember that we implicitly operated in
both, the $p,q$ and $a,a^\dagger$ basis, which are related in a non trivial
way if divided powers algebras are concerned. A more careful exposition will 
be presented elsewhere.

Before moving to quantum field theory in general, it should be borne in
mind that the first two examples, symmetric function theory, and quantum
oscillators, are closely related via two-dimensional quantum
field theory and the fermion-boson correspondence \cite{kac:1998a}. 
From the line of development we have followed, the symbolic role played
by vertex operators (fields), as exponentials of oscillator modes 
\cite{jarvis:jung:1992a,jarvis:yung:1993a,salam:wybourne:1992a}, 
could be expected to be intimately related to that of plethysms 
\cite{carre:thibon:1992a,backer:1996a} as functors at a categorial level in 
symmetric function theory \cite{macdonald:1979a}. 

\subsection{\label{sec:Renormalization}Combinatorics of renormalization 
in quantum field theory}

In this paragraph we exhibit the implications of the Dirichlet Hopf algebra in
the theory of renormalization of quantum fields. We concentrate on the
combinatorial side, but want to emphasize that due to the categorial structure
of our considerations we expect the scheme to be generally valid. A functorial
description supporting this view will be given elsewhere.

Our mathematical approach makes contact with renormalization through the work of
Brouder and Schmitt \cite{brouder:schmitt:2002a}. A Hopf algebraic formulation
of pQFT was given in \cite{brouder:fauser:frabetti:oeckl:2002a}. The arena of 
renormalization is the bialgebra $B$ of normal ordered (scalar) fields 
$\phi^n(x)$. Here $x\in\openR^{1,3}$ is a space-time point\footnote{%
Strictly speaking we need to consider test functions sufficiently smooth
and localized, but that does not matter for the present algebraic discussion.
So we use the loose notation.}
and $n\in\ZP$. The algebra and coalgebra structures as used in physics are 
given as 
\begin{align}
\phi^n \phi^m 
&= \phi^{n+m} \nn
\delta_B \phi^n 
&= \sum_{k=0}^n {n\choose k} \phi^k\otimes \phi^{n-k} \nn
\epsilon_B(\phi^n) &= \delta_{n,0}  
\end{align}
This pair of operations thus comprises addition and the unrenormalized 
coproduct of addition. The choice reflects the fact that the dual 
$\phi^{\#}(x)$ is assumed to obey a divided powers algebra structure and the 
coproduct is the dualized product of the divided powers algebra of the dual 
fields. Then one defines the symmetric algebra $\sym[B]$ over the 
\textit{module} underlying $B$. A monom in $\sym[B]$ reads
\begin{align}\label{aphi}
a\,=\, :\phi^{n_1}(x_1)\ldots\phi^{n_k}(x_k):
&= (\phi^1)^{(n_1)}(\phi^2)^{(n_2)}\ldots (\phi^k)^{(n_k)}
\end{align}
where the right hand side is rewritten in the same type of notation as we
employed for symmetric functions, compare eqn. (\ref{monomialM}), only the
`letter' of our alphabet, generating the algebra, is now the quantum field
$\phi$. The coproduct induced from $B$ is given as
\begin{align}
\delta_B \Big( (\phi^1)^{(n_1)}\ldots(\phi^k)^{(n_k)} \Big)
&=
\sum_{i_1}^{n_1}\ldots\sum_{i_k}^{n_k}
{n_1\choose i_1}\ldots{n_k\choose i_k} \nn
&~~~~~~
(\phi^1)^{(i_1)}\ldots(\phi^k)^{(i_k)}\otimes
(\phi^1)^{(n_1-i_1)}\ldots(\phi^k)^{(n_k-i_k)}
\end{align}
these are \textbf{normal ordered fields}. The time ordered product of such
a monomial in the field $\phi$ is given \cite{brouder:fauser:frabetti:oeckl:2002a}
in analogy with the treatment of Epstein-Glaser, as
\begin{align}
T(a) &= \sum t(a_{(1)}) a_{(2)}
\end{align}
where $t : \sym[B^+] \rightarrow \openk$ is an appropriate $1$-cochain. Note 
that this is a branching operator as defined above in sections \ref{sec:2-3}, 
\ref{sec:2-4}, \ref{sec:3-4} and \ref{sec:3-5}, explicitly reading
$T = \cdot(t\otimes)\delta_B$. Such operators play a fundamental 
role in the theory of group branchings 
\cite{fauser:jarvis:2003a,fauser:jarvis:king:2005a} from
where we borrow their name. The \textbf{Wick theorem} \cite{fauser:2001b} 
takes the form 
\begin{align}
T(: (\phi^1)^{(n_1)}\ldots(\phi^k)^{(n_k)}:)
&=
\sum_{i_1}^{n_1}\ldots\sum_{i_k}^{n_k}
{n_1\choose i_1}\ldots{n_k\choose i_k} \nn
&~~~~~~
t(:(\phi^1)^{(i_1)}\ldots(\phi^k)^{(i_k)}:)
:(\phi^1)^{(n_1-i_1)}\ldots(\phi^k)^{(n_k-i_k)}:
\end{align}
in accordance with
\cite{fauser:2001b,brunetti:fredenhagen:2000a,epstein:glaser:1973a} and 
the development in \cite{brouder:schmitt:2002a}. In physics the renormalization 
is then introduced via a second time ordered product $\tilde{T}$ as follows. For 
$a$ as in eqn. (\ref{aphi}) one defined the $b^i$ are the nonempty parts 
of $a$. Define further a map $O$ as
\begin{align}
\label{ttilde}
O : \sym[B] &\rightarrow B\nn
\tilde{T}(a) &= \sum_\lambda T( :O(b^1)\ldots O(b^l):)
\end{align}
where $\lambda$ is a setpartition of $\sum_i n_i$ into $l$ nonempty 
parts. The operator $O$, encoding an renormalization scheme e.g. extracting pole
parts, was introduced in Bogoliubov-Shirkov \cite{bogoliubov:shirkov:1980a} 
(under a different name). It is convenient to set $\tilde{T}(1)=O(1)=1$ and for 
$a\in B$ set $\tilde{T}(a)=a$, hence one gets $O(a)=a$. Using the standard recursion,
employing the proper cut coproduct, one comes up with
\begin{align}
\tilde{T}(a) 
&=T(a)+O(a)+{\sum_\lambda}^\prime 
T(:O(b^1)\ldots O(b^l):)
\end{align}
Brouder and Schmitt show that $O$ is also a branching operator, and can be
defined using a 1-cochain $c(a)=\epsilon_B(O(a))$ as\footnote{%
The first equality can be proved for logarithmic divergences, while in the
general case it may look different [private communication Ch. Brouder].
}
\begin{align}
O(a)&= \sum c(a_{(1)}) a_{(2)}\nn
c(a)&=\tilde{t}(a)-t(a)
 -\sum^\prime_\lambda \sum c(b^1_{(1)})\ldots c(b^l_{(1)})
t(:c(b^1_{(2)})\ldots c(b^l_{(2)}:))
\end{align}
Formally one obtains a new 1-cochain $c=\tilde{t}-t$ allowing to write the
abstract form of the branching operators $O$ as
\begin{align}
\label{o}
O(a) &= \sum c(a_{(1)}) \prod a_{(2)}
\end{align}
The relation to renormalization is now given by combining eqn. (\ref{ttilde})
and (\ref{o}) to obtain
\begin{align}
\tilde{T}(a) 
&= \sum T(:O(b^1)\ldots O(b^l):)\nn
&= \sum^\prime_\lambda \sum c(b^1_{(1)})\ldots c(b^l_{(1)})
   T(:\prod b^1_{(1)}, \ldots, \prod b^l_{(2)}) \nn
&= \sum C(a_{[1]}) T(a_{[2]}) 
\end{align}
where this is related to the coproduct in $\sym\sym[B]^+$. It is remarkable
that the renormalization map is given by $C(a)=c(a)$ for $a\in \sym[B]^+$ 
and $C(uv)=C(u)C(v)$ for $u,v\in \sym\sym[B]^+$. This is exactly the form of
maps obtained by the branchings induced by plethysms as obtained in 
\cite{fauser:jarvis:king:wybourne:2004a}. This formula, via some steps of
identification, carried out in \cite{brouder:schmitt:2002a}, finally makes
contact to the Epstein-Glaser framework of renormalization. 

While our treatment remained formal and followed closely the approach of
Brouder and Schmitt, we hope to have shown that the lift from the coproduct of
addition to the coproduct of multiplication (causing the group like coproduct
action on $a$ indicated by the Sweedler indices $a_{[i]}$) is reminiscent of
the same process as lifting addition to multiplication. We denoted this
mnemonically as $\DM=\delta^\DP$. A similar process can be found in group
theory and describes the branchings of characters of subgroups of $GL(n)$
which fix a tensor of arbitrary Young symmetry type $\pi$ 
\cite{fauser:jarvis:king:wybourne:2004a}. The insight which can be drawn 
from this work is, that branchings in group theory and the reorderings induced 
by branching operators, as demonstrated in \cite{fauser:2001b}, are in general
based on 1-cochains which are \textbf{not product homomorphisms}. In the 
case of addition and multiplication it was seen that many number theoretic 
functions are not complete multiplicative. The cure was to introduce an 
unrenormalized (binomial) coproduct and to invent a subtraction scheme to
remove the superfluous terms which were introduced to establish the
homomorphism property at the end of a calculation. Our discussion sheds some 
light on our naming scheme of the arithmetic coproducts and pairings. Actually
the non Hopf convolutions $(+,\DP)$ and $(\cdot,\DM)$ are of mathematical
and physical interest. However they have bad algebraic behaviour forming a Hopf
gebra only, so that it is more natural to adopt another coproduct
and/or pairing map imposing a nice algebraic behaviour (Hopf algebra and
homomorphism property). To get rid of the artificially introduced terms one
invents then a subtraction scheme, called `renormalization', which extracts
the searched for results. We named our coproducts and pairings so that
they comply with their potential usage in renormalization theory in pQFT. 

\section{Concluding remarks}
\setcounter{equation}{0}\setcounter{mycnt}{0}

It is our hope, to have convinced the reader, that the intimately linked
structures of addition and multiplication also have a perfectly justified
dual life. The coproducts of addition and multiplication play important roles
in various branches of mathematics. It was the impetus of this work to make
this obvious. The relation between addition and multiplication found its
mnemonic counterpart in the formula $\DM=\delta^\DP$.

The subtle distinction between multiplicative and complete multiplicative
functions in arithmetic number theory has an algebraic counterpart in the fact
that the additive and multiplicative convolutions are antipodal but 
\textbf{fail to be Hopf algebras}. This deficiency may be seen as the
major source of hard problems in (analytic) arithmetic number theory. Since
numbers can be addressed in various ways it is difficult to disentangle the
role they perform in various settings. Numbers may just count something 
(states in physics). Numbers can operate on other numbers (operators in 
physics). Numbers may act as linear forms (dual states) etc. Our
investigations have shown that, if a careful identification is done, one
obtains new insight into the machinery of symmetric functions, quantum 
mechanics and quantum field theory. This has some implications for number
theory also, since it implies that such methods as `renormalization', after 
being turned into proper mathematics, may help to solve some long standing
problems.

In this work `renormalization' was formalized as the following procedure:

{\bf a:} Observe that the positive selfadjoint convolutions of addition 
and multiplication, where products and coproducts are related by Kronecker 
duality, are antipodal convolutions but fail to be Hopf and form a
\textbf{Hopf gebra} only \cite{fauser:2002c,fauser:oziewicz:2001a}.
This stage is missing in physics, where the modelling is done using the
mathematical structures appearing in stage {\bf b:}, e.g. in number theory
we need an `unrenormalization map' to proceed to step {\bf b:}.

{\bf b:} Introduce a second duality by imposing a new coproduct which is
identical on primitive elements to the dualized one but extended as an
algebra homomorphisms. This new coproduct forms with the related product
by construction a Hopf algebra with all the nice algebraic properties.

{\bf c:} Compute via Laplace expansion \cite{grosshans:rota:stein:1987a,%
rota:stein:1994a,fauser:2002c,brouder:fauser:frabetti:oeckl:2002a}, which 
is available due to construction of the unrenormalized coproducts, a new 
unrenormalized pairing. It is this pairing which reflects the algebraic 
relation between variables and derivations, which is typical for the
description of quantum systems.

{\bf d:} Invent a \textbf{subtraction scheme}, called `renormalization',
which extracts those terms in the Hopf algebraic convolution which have
been introduced by the unrenormalized coproduct to obtain the nice algebra. 
Prove that the Hopf algebra together with the subtraction scheme gives
the same results as the original antipodal convolution (Hopf gebra).  
This subtraction scheme can be employed via a deformation or a 
Rota-Baxter operator, as we demonstrated in our quantum mechanical
example in section \ref{sec:Occupation}.

To solidify our point of view we have applied our framework to number
theory, symmetric function theory etc. An amazing fact is further,
that our treatment of renormalization benefits from a number theoretic 
approach, and a characteristic free approach. In this sense it is not
a miracle that number theoretic functions such as multiple zeta-values
and Clausen functions appear in the evaluation of a renormalized quantum
field theory. The idea, proposed by Cartier \cite {cartier:2001a}, 
of a motivic Galois group seems hence to be related to the theory of
Witt vectors (sec. \ref{sec:Witt}) and the Witt functor, which actually
provided us in the theory of symmetric functions with the correct basis. 

This paper also provided some starting points for a characteristic free
approach to quantum mechanics. The appearance of complex numbers, an 
algebraically closed field, is often argued to be a key feature, e.g. for
producing interference effects, but we doubt this. A `phase' may be modelled
by a finite cyclic group also. The expectation values can be obtained in a
topos theoretic setting using more general truth objects and therewith
related subobject classifiers. This will be explored elsewhere. We think
that the present work shows at least, that for the identification of the
algebraic structures involved in quantum mechanics and quantum field theory, 
a characteristic free approach to quantum mechanics would be of great help.
Especially the interpretation of combinatorial factors, normalizations etc. 
would benefit from such a view, even if the complex number field is finally
adopted. The appearance of number theoretic functions in renormalization
supports this point of view.,

While the present paper provided a detailed exposition of the Dirichlet
Hopf algebra, it is necessary to point out, what should be done to unveil 
the underlying categorial structure. A careful study of branching operators
is necessary. These operators not only appear in the theory of group
branchings \cite{fauser:jarvis:2003a}, but emerge also in the treatment of
random walks \cite{ellinas:tsohantjis:2001a} and in the study of entanglement
of quantum states. The is also a close relation to categorial logic. We 
noticed, that the Hopf algebra cohomology becomes delicate if the 
homomorphism axiom no longer holds, even if it is not explicitly needed in
the proof of a particular statement. Hence this cohomology has to be revisited,
to show where the homomorphism axiom enters, and what can be still obtained
having only the weakened multiplicativity axiom. The failure of the
homomorphism axiom (complete multiplicativity) provides the key feature for
deducing new branching rules and character formulae for non semisimple
subgroups of $gl(n)$, as is shown in 
\cite{fauser:jarvis:king:wybourne:2004a,fauser:jarvis:king:2005a}.
The related problem of computing plethysms is a long standing one and
algorithmically not solved in a satisfactory way. An analysis along the
same lines as of the present paper is in progress, and a Hopf algebraic
version of the plethysm may help to develop better algorithms if the coalgebra
structure is at hand also. This is ongoing work \cite{fauser:jarvis:2005a},
but recall section \ref{sec:Witt} on Witt vectors.

Of course loose ends are left with renormalization. Most of them need an
identification of the proper algebraic structures involved, and
combinatorics is a beautiful help in doing so. Indeed, in our oppinion
one needs to step back and to describe the categorical backbone of the
structure, for more efficient identification of the seperate algebras
involved. In the modelling of quantum field theory algebraic structures are
mostly used without semantic or syntactic distinctions. Such a categorial
picture will naturally include 2-categories: we have already used 
$\pleth[A]$ and $\sym\sym[B]^+$ whose underlying modules form 2-vector 
space structures. In terms of operations, multiplication is the iterated
addition, addition is the iterated successor map of Peano. In this sense, 
`renormalization' seems to be necessary to interlock those iterated
structures properly. To understand this process in a mathematically
rigorous way is therefore of great importance. 

It would hence be an interesting task to figure out the proper identification
of the Bogoliubov map $C$ back in the theory of symmetric functions. Actually
this has to be a basis transformation between the monomial symmetric basis,
which we used for the construction of $\sym\sym[B]^+$ and a basis of the
form 
\begin{align}
\tilde{T}(a) 
&= C(a_{[1]}) T(a_{[2]}) \nn
&= {\sum_\lambda}^\prime \sum c(b^1_{(1)})\ldots c(b^l_{(1)})
   T(:\prod b^1_{(1)}, \ldots, \prod b^l_{(2)}) \nn 
&= {\sum_\lambda}^\prime \sum (b)^{(1)}\ldots (b)^{(1)}
   T(:\prod b^1_{(1)}, \ldots, \prod b^l_{(2)}) \nn
\end{align}
where the prefactors are evaluated under the Gessel map at monomials
of type
\begin{align}
h_\lambda &= (a^{\#})^{(\lambda_1)}\circ\ldots(a^{\#})^{(\lambda_n)}
\end{align}
as can be guessed from \cite{rota:stein:1994b} page 13063, in-line formula
before proposition 1 (notions also taken from those sources). Rota and Stein
identify the $h_\lambda$ in a dual setting, therefore our usage of $a^{\#}$.
Actually one expects these functions to be related to the $r_\lambda$ basis
and its dual $q_\lambda$ of the Witt ring. It is however very difficult
to match these objects since common treatments do not take care of the
proper algebraic distinction between $a$, $a^{\#}$ and the various
algebraic structures which arise from the $\pleth[A]$ Hopf algebra and
its dual $\textrm{Brace}\{A\}$, see loc. cit.

Our list of further opportunities and open problems could be prolonged
considerably, and the attentive reader will have noticed these during the
course of reading.

\vskip 2ex
\noindent{\bf Acknowledgement:} The first author gratefully acknowledges
financial support from the ARC (project number DP0208808) and of the 
Alexander von Humboldt Foundation for travel support in the `sur place' 
program. It is also a pleasure to thank the School of Mathematics and
Physics of the University of Tasmania at Hobart where this work was
done for their hospitality. PDJ thanks the Max Planck Institut f\"ur
Mathematik in den Naturwissenschaften, Leipzig, and the
Alexander von Humboldt Stiftung, for support.

\begin{appendix}

\section{Matrix analogy for addition}
\setcounter{equation}{0}\setcounter{mycnt}{0}

In \cite{fauser:2004a} we investigated some facts about positive selfadjoint
Hopf algebras using singular value decomposition (SVD) methods. We will use
this device here too, to establish a few relations about the product and 
coproduct of addition. First we introduce the rectangular `multiplication
table' $m$ of addition
\begin{align}
\begin{array}{c|ccccccccccc}
+ & (0,0) & (1,0)&(0,1) & (2,0)&(1,1)&(0,2) & (3,0)&(2,1)&(1,2)&(0,3) & \ldots 
\\\hline
0 & 1 & 0 & 0 & 0 & 0 & 0 & 0 & 0 & 0 & 0 & \ldots \\ 
1 & 0 & 1 & 1 & 0 & 0 & 0 & 0 & 0 & 0 & 0 & \ldots \\ 
2 & 0 & 0 & 0 & 1 & 1 & 1 & 0 & 0 & 0 & 0 & \ldots \\ 
3 & 0 & 0 & 0 & 0 & 0 & 0 & 1 & 1 & 1 & 1 & \ldots \\
\vdots & \vdots &&&&&&&&& \vdots & \ddots
\end{array}
\end{align}
This table is infinite and reads like this: The addition of the pair
in the first row gives the sum (one term in fact) of the first column
weighted by the column below the pair. E.g. $+(2,1)= 0\cdot 0+
0\cdot 1+0\cdot 2+ 1\cdot 3 + \ldots$. The Kronecker addition coproduct
has as its comultiplication table $m^T$, the 
\textbf{section coefficients}, the transposed infinite matrix.

Using ideas from SVD we may form the matrices $A=m \,m^T$ and $B= m^T\, m$
both rectangular and symmetric by construction. $A$ and $B$ have up to a
kernel the same eigenvalues, which are squares of the singular values attached
to $m$ and $m^T$. We compute
\begin{align}
A&=
\left[
\begin{array}{ccccc}
1 & 0 & 0 & 0 & \ldots \\
0 & 2 & 0 & 0 & \ldots \\
0 & 0 & 3 & 0 & \ldots \\
0 & 0 & 0 & 4 & \ldots \\
\vdots & & &\vdots & \ddots
\end{array}
\right] \nn
B&=
\left[
\begin{array}{cccccccccccccccc}
1 & 0 & 0 & 0 & 0 & 0 & 0&0&0&0 &\ldots \\
0 & 1 & 1 & 0 & 0 & 0 & 0&0&0&0 &\ldots \\
0 & 1 & 1 & 0 & 0 & 0 & 0&0&0&0 &\ldots \\
0 & 0 & 0 & 1 & 1 & 1 & 0&0&0&0 &\ldots \\
0 & 0 & 0 & 1 & 1 & 1 & 0&0&0&0 &\ldots \\
0 & 0 & 0 & 1 & 1 & 1 & 0&0&0&0 &\ldots \\
0 & 0 & 0 & 0 & 0 & 0 & 1&1&1&1 &\ldots \\
0 & 0 & 0 & 0 & 0 & 0 & 1&1&1&1 &\ldots \\
0 & 0 & 0 & 0 & 0 & 0 & 1&1&1&1 &\ldots \\
0 & 0 & 0 & 0 & 0 & 0 & 1&1&1&1 &\ldots \\
\vdots & & & & & & & & & \vdots & \ddots
\end{array} 
\right]\nn
\end{align} 
It is easily checked that $B$ has also eigenvalues $1,2,3,4,\ldots$
and an infinite kernel. Hence one sees, that the addition has singular values
$\sqrt{1},\sqrt{2},\sqrt{3},\ldots$ which do \textbf{not} belong to $\ZP$.

From the above cited paper \textit{Fauser, loc. cit.} we know, that the 
following two operators have then up to the kernel the same spectrum
\begin{align}
(+\circ \DP)(n) = n_{(1)}+n_{(2)} &= (n+1)\, n \nn
 \prod_{k\ge 0} ((+\circ \DP) - (k+1)) &=0 \nn  
 (\DP\circ +)^\infty\, \prod_{k\ge 0} ((\DP\circ +) - (k+1)) &=0  
\end{align} 
The operator in the last line is, however, not diagonal. Eigenvectors are
provided by the Kronecker coproduct of addition due to construction,
see \cite{fauser:2004a}.

\section{Matrix analogy for multiplication}
\setcounter{equation}{0}\setcounter{mycnt}{0}

In the same way as we did with addition, we proceed with multiplication and
compute the first elements of the multiplication table $m$
\begin{align}
\begin{array}{c|ccccccccccc}
\cdot & (1,1) & (2,1)&(1,2) & (3,1)&(1,3) & (4,1)&(2,2)&(1,4) &(5,1)&(1,5) & \ldots 
\\\hline
1 & 1 & 0 & 0 & 0 & 0 & 0 & 0 & 0 & 0 & 0 & \ldots \\ 
2 & 0 & 1 & 1 & 0 & 0 & 0 & 0 & 0 & 0 & 0 & \ldots \\ 
3 & 0 & 0 & 0 & 1 & 1 & 0 & 0 & 0 & 0 & 0 & \ldots \\ 
4 & 0 & 0 & 0 & 0 & 0 & 1 & 1 & 1 & 0 & 0 & \ldots \\
5 & 0 & 0 & 0 & 0 & 0 & 0 & 0 & 0 & 1 & 1 & \ldots \\
\vdots & \vdots &&&&&&&&& \vdots & \ddots
\end{array}
\end{align}
Note the different indexing of the first row by pairs of divisors of $n$.
Once more, the section coefficients of the coproduct obtained by Kronecker
dualization $\DM=m^T$ is given by the transposition. In the SVD style, we 
can once more form the symmetric matrices $A=m \,m^T$ and $B= m^T\, m$,
which read
\begin{align}
A&=
\left[
\begin{array}{cccccc}
1 & 0 & 0 & 0 & 0 & \ldots \\
0 & 2 & 0 & 0 & 0 & \ldots \\
0 & 0 & 2 & 0 & 0 & \ldots \\
0 & 0 & 0 & 3 & 0 & \ldots \\
0 & 0 & 0 & 0 & 2 & \ldots \\
\vdots & & & & \vdots & \ddots
\end{array}
\right] \nn
B&=
\left[
\begin{array}{cccccccccccccccc}
1 & 0 & 0 & 0 & 0 & 0 & 0 & 0 &0&0 &\ldots \\
0 & 1 & 1 & 0 & 0 & 0 & 0 & 0 &0&0 &\ldots \\
0 & 1 & 1 & 0 & 0 & 0 & 0 & 0 &0&0 &\ldots \\
0 & 0 & 0 & 1 & 1 & 0 & 0 & 0 &0&0 &\ldots \\
0 & 0 & 0 & 1 & 1 & 0 & 0 & 0 &0&0 &\ldots \\
0 & 0 & 0 & 0 & 0 & 1 & 1 & 1 &0&0 &\ldots \\
0 & 0 & 0 & 0 & 0 & 1 & 1 & 1 &0&0 &\ldots \\
0 & 0 & 0 & 0 & 0 & 1 & 1 & 1 &0&0 &\ldots \\
0 & 0 & 0 & 0 & 0 & 0 & 0 & 0 &1&1 &\ldots \\
0 & 0 & 0 & 0 & 0 & 0 & 0 & 0 &1&1 &\ldots \\
\vdots & & & & & & & & & \vdots & \ddots
\end{array} 
\right]\nn
\end{align} 
From this description one sees that the linearly ordered basis is a
particular poor choice for ordering the involved basis elements. The monoid 
$\ZP$ with adjoined multiplication map should be considered as a poset under
the partial order of mutual divisibility. Such a poset is graded by the 
numbers of prime factors of a number. The above matrix $A$ would then come
up with an infinite list of eigenvalues $1,2^\infty,3^\infty,4^\infty,5^\infty,\ldots$ where
$\infty$ notifies the countable infinity of primes. The same would be true
for the matrix $B$, since it has the same eigenvalue structure as $A$ up to
a kernel due to SVD theory. The characteristic polynomial has hence the
primes as roots and all of them have an infinite degeneracy
\begin{align}
A &:\qquad \Pi_i (\cdot\circ\DM -p_i)^\infty =0 \nn
B &:\qquad (\cdot\circ\DM)^\infty\,\Pi_i (\cdot\circ\DM -p_i)^\infty =0,
\end{align}
where the products run over all primes. Of course, these formulae make only
formal sense due to the infinite degeneracy.
\end{appendix}

\addcontentsline{toc}{section}{References}
\small{

}
\end{document}